\documentclass[preprint]{aastex}

\newcommand{\degsq}{deg$^2$}

\shortauthors{Harding et al.}
\shorttitle{Observations of Tidal Streams}

\begin{document}

\title{Mapping the Galactic Halo III. 
  Simulated Observations of Tidal Streams}

\author{Paul Harding} 
\affil{Steward Observatory, University of
  Arizona, Tucson, Arizona 85726 \\ 
  electronic mail:harding@billabong.astr.cwru.edu} 
\author{Heather L.Morrison
\footnote{Cottrell Scholar of Research Corporation and NSF
    CAREER fellow}} 
\affil{Department of Astronomy\footnote{and Department of Physics},
  Case Western Reserve University, Cleveland
  OH 44106-7215 \\ electronic mail: heather@vegemite.astr.cwru.edu}
\author{Edward W. Olszewski} 
\affil{Steward Observatory, University of
  Arizona, Tucson,
  AZ 85721\\
  electronic mail: edo@as.arizona.edu} 
\author{John Arabadjis, Mario Mateo and R.C. Dohm-Palmer}
\affil{Department of Astronomy, University of Michigan, 821
  Dennison Bldg., Ann Arbor, MI 48109--1090\\
  electronic mail: jsa@space.mit.edu, mateo@astro.lsa.umich.edu, rdpalmer@astro.lsa.umich.edu} 
\author{Kenneth C.
  Freeman and John E. Norris} \affil{Research
  School of Astronomy \& Astrophysics, The Australian National University,
  Private Bag,
  Weston Creek PO, 2611 Canberra, ACT, Australia\\
  electronic mail: kcf@mso.anu.edu.au,jen@mso.anu.edu.au}


\begin{abstract}
  We have simulated the evolution of tidal debris in the Galactic halo
  in order to guide our ongoing survey to determine the fraction of
  halo mass accreted via satellite infall.  Contrary to naive
  expectations that the satellite debris will produce a single narrow
  velocity peak on a smooth distribution, there are many different
  signatures of substructure, including multiple peaks and broad but
  asymmetrical velocity distributions. Observations of the simulations
  show that there is a high probability of detecting the presence of
  tidal debris with a pencil beam survey of 100 square degrees.  In
  the limiting case of a single $10^{7}$ $M_{\odot}$ satellite
  contributing 1\% of the luminous halo mass the detection probability
  is a few percent using just the velocities of 100 halo stars in a
  single 1 \degsq\ field. The detection probabilities scale with the
  accreted fraction of the halo and the number of fields surveyed.
  There is also surprisingly little dependence of the detection
  probabilities on the time since the satellite became tidally
  disrupted, or on the initial orbit of the satellite, except for the
  time spent in the survey volume.
\end{abstract}

\keywords{Galaxy: halo --- Galaxy: formation --- Galaxy: kinematics and dynamics}

\section{Introduction}
Studies of our Galaxy's halo have an important role to play in
understanding the process of galaxy formation.  Classical scenarios of
halo formation such as that of \citet{els} have given way to a more
mature view of galaxy formation in the context of the formation of
structure in the universe \citep{ms94,sdm2000}. These
ideas were foreshadowed by \citet{sz78} from an observational perspective.
Hierarchical structure formation models now have the resolution that
allows them to make predictions on scales as small as the Local Group.
There are some puzzling early results: the recent simulations of
\citet{klypin99a} and \citet{moore99a} predict that there should be
many more dwarf satellites at $z=0$ in the Local Group than are
currently seen (if the dark halos in their simulations can be
associated with dwarf galaxies). Were these satellites torn apart to
form the stellar halo? If so, some of them should still be visible as
tidal streams. In fact, we now have strong evidence that the halo was
formed at least in part by the recent accretion of one or more
satellites. Earlier papers in this series \citep{spag1,spag2} have reviewed the
evidence from both kinematics and star counts for substructure in the halo. 
It is clear that further studies of the halo of the Milky
Way and its satellites are needed to clarify these discrepancies.

To date, theoretical investigations have focused on the accretion and
destruction of a limited range of satellite orbits. In this paper we
concentrate on estimating the detectability of tidal debris from a
large range of initial satellite orbits --- in other words, our focus
is more statistical and observational.  Our aim is to estimate the
probability of detecting kinematic substructure using currently
feasible observational strategies.  We hope to provide a bridge for
observers from more theoretical papers on satellite destruction and
phase mixing \citep{helmi99,tre_geom} to find the optimum
observational techniques to detect kinematic substructure in the halo.

Dynamical models are not only helpful in tracing the history of the
debris, but can also be used to plan the survey strategy and to help
interpret results. For example, the early detection of kinematic
substructure at the NGP by \citet{srm94} was puzzling because of its
large velocity spread ($\sigma \simeq 100 $ km/s in each component of
the proper motions). This is now understandable in terms of multiple
wraps of a single orbit \citep{helmihip}.  A more recent puzzle is the
discovery of the Sloan over-density along 30 degrees of their
equatorial strip \citep{sloanrr,yanny2000}. Was this just a fortunate
coincidence that a narrow tidal feature was aligned with the celestial
equator, or is the feature in fact more spatially extended?

In this paper we assemble the tools for using our
survey data to measure the fraction of the halo that has been
accreted.  In Section 2 we begin by outlining the procedure used to
create the Galaxy model, satellite model and tidal streams.  Section 3
shows the results from six satellite orbits to highlight properties of
the spatial and velocity evolution of the debris. We then illustrate
how these results map into observable coordinates.  In section 4 we
discuss how we ``observe'' the models to determine their detection
probability using our survey strategy. We then discuss the factors
that determine the detection of the debris on different orbits at
different times.  Section 5 looks at how sensitive the detection
probabilities are to specific observing procedures. We note how the
properties of currently available instrumentation figure into these
constraints.  Section 6 discusses implications of these results for
detecting substructure in the halo, in particular which new methods
need to be developed to efficiently trace detected substructure.

\section{Simulating the Galaxy's Halo} 

We have created model stellar halos from a mix of a ``lumpy'' and
smooth components. The lumpy component represents debris from
satellite accretions which still retains information about its
origin.  The smooth component represents the portion of the halo which
is so well-mixed that its velocity distribution can be approximated by
a velocity ellipsoid.  We created the lumpy component by evolving
satellites on a range of initial orbits in the Galaxy's potential.
(We describe in Section 2.2 our library of initial satellite orbits.)
We then sample from the remains of the satellites at various times
during their destruction.  Two versions of the smooth component have
been created.  One has a radially extended velocity ellipsoid similar
to the underlying distribution from which the satellite orbits were
chosen.  The second used an isotropic velocity ellipsoid.

Many realizations of the model halos need to be created so that we can
become familiar with the range of accretion signatures likely to be
present in our observational data. We wish also to quantify the
detection probability as a function of the initial satellite orbit,
its time since destruction, the galactic coordinates $(l,b)$ of the
fields surveyed, and how the detection probabilities scale with the
accreted fraction of the halo.

In order to maximize the number of different satellite orbits
available to select from in the creation of our model halos, we have
made a number of simplifications to reduce the computation time
required to a reasonable level. We have used a fixed potential for the
Galaxy and have neglected the self-gravity of the satellites as we
destroy them to make tidal debris.  As a general principle we have
attempted to match the observed properties of the Galactic halo
wherever possible in our choice of parameters.  Details of the models
are given in the following sections.

\subsection{The Potential}
We have followed the prescription of \citet{jsh95} for the Milky Way's
potential, which provides a good match to the rotation curve of the
Galaxy. The potential consists of three components.
 The disk is described
by a Miyamoto-Nagai potential \citep{mn75},
\begin{eqnarray*}   \Phi_{\rm disk}=-{GM_{\rm disk} \over
                 \sqrt{R^{2}+(a+\sqrt{z^{2}+b^{2}})^{2}}} ;
\end{eqnarray*} 

the central stellar density of the bulge/bar and inner halo is
represented by a \citet{lh90} potential
        \[ \Phi_{\rm spher}=-{GM_{\rm spher} \over r+c} ; \] 
        and the dark halo by a logarithmic potential:
        \[ \Phi_{\rm halo}=v_{\rm halo}^2 \ln (r^{2}+d^{2}) . \]
        
        Here $M_{\rm disk}=1.0 \times 10^{11} M_{\odot}, M_{\rm
          spher}=3.4 \times 10^{10}M_{\odot}, v_{\rm halo}= 128 $
        km/s, and lengths $ a=6.5$ kpc,$ b=0.26$ kpc, $c=0.7$ kpc, and
        $d=12.0$ kpc.  The adoption of a fixed potential should not
        have a significant influence on our results, since the mass of
        the stellar halo \citep[$ \sim 10^{9} M_{\odot}, $][]{hlm96}
        is only a small fraction of the total mass of the Galaxy  \citep[$
        \sim 10^{12} M_{\odot} $, ][]{dz89}.
        
        The $\Lambda$-dominated cosmology predicts that most of
        the Galaxy's mass was assembled in the first 5 Gyr (in
        contrast to the $\Omega=1$ CDM cosmology \citep{nfw95} where
        assembly happens later), and therefore a fixed potential is a
        good approximation for the next 10 Gyr.  The growth of the
        Galaxy's potential, provided it happens on time-scales longer
        than the satellite orbital period should not be significant in
        changing the outcome of these simulations \citep{hz99}.

\subsection{Orbit selection}


The Galaxy's visible halo is extremely centrally concentrated, with
density $\rho \propto r^{-3}$ or even $r^{-3.5}$
\citep{abi85,zin93,gwp94,spag1}.  \citet{yanny2000} have also shown
that, if the excess of BHB stars associated with the Sloan stream are
excluded, then the density of BHB stars in their survey falls as
$r^{-3.1}$.  To create a density distribution this steep there needs to
be a significant fraction of orbits with small mean radii. It is
impossible to make a halo steeper than $r^{-3}$ with only radial
orbits with large apocenters with this potential, since the time spent
traversing a segment of the orbit always decreases faster than the
volume element at that radius.

First we constructed an equilibrium distribution of orbits in the
Galaxy's potential. They were chosen to have an $r^{-3.0}$ density
distribution, and radially anisotropic velocity ellipsoid of
$(\sigma_{r}, \sigma_{\phi}, \sigma_{\theta} )$ = (150,110,100) km/s,
similar to the values determined by \citet{cy98}. This
selection\footnote{This procedure was used so that in future we can
consistently create model halos with a varying mix of smooth and
accreted components. However, in this paper will concentrate on models
with only a single accreted satellite.}  produces a range of orbital
energies and angular momenta allowed by the potential, weighted
towards more radial orbits.  The satellite orbits have a distribution
of eccentricities similar to those predicted by the high resolution
structure formation models \citep{ghigna98,bosch99}.

A subset of 180 orbits were sampled from this distribution, to have
pericenters between 0.4 and 26 kpc and mean radii \footnote{The time
weighted mean radius.} greater than 8 kpc.
These orbits will serve as center-of-mass orbits for the satellites
whose destruction we will study below.  The properties of the selected
orbits are summarized in Figure \ref{orbit_properties}. In the
following sections we will refer to orbits by their mean radius. (The
approximate apocenter in kpc and orbital period in Gyr can be
estimated by scaling the mean radius of the orbit by 1.4 and 0.02
respectively.)

These selection criteria allow us to focus our computational resources
on the orbits of interest without introducing any significant
statistical biases. Orbits with pericenter less than 0.5 kpc disperse
in phase space within a few orbits. Thus they are more appropriately
treated as part of the smooth component of the halo.  Orbits with
pericenter greater than 27 kpc can never fall into our survey volume.
Our survey has a magnitude limit of V $<$ 20 \citep{spag1}, which
translates into a maximum distance of $R_{\odot} < 20$ kpc for dwarf
stars near the turnoff. Dwarfs are the only halo tracer population
where it is possible to efficiently obtain a sample of $\simeq$ 100
velocities within a small area of the sky, due to the rarity of
luminous halo stars \citep{spag1}. Turnoff stars are the most luminous
tracers for which this is possible.  Orbits with mean radii less than
8 kpc, whatever their pericenter, are also assumed to be part of the
smooth component of the models.  They were presumably accreted early
in the Galaxy's history and their kinematics will have lost any
measurable information about their origin due to phase mixing. Early
on, violent relaxation may also have contributed to the loss of
information.

How can a satellite be accreted from outside the Milky Way and have a
very small mean radius? This is only possible via dynamical friction.
However, recently accreted satellites will not have had time to sink
to small mean radii --- the time-scale for dynamical friction is too
long \citep{dynf99}. Only satellites more massive than $10^{11}
M_\odot$ would decay sufficiently rapidly to have mean radii 10 kpc or
less.  However, satellites this massive, if they remained
predominantly intact, would cause significant heating of the old thin
disk, which is not observed \citep{walker96,edv93}.

\subsection{Evolution of satellites into tidal streams}

The satellites representing small dwarf galaxies were created using a
tidally truncated Plummer model \citep{bat87} of $10^{7} M_{\odot}$,
populated with 200,000 particles.  The satellites are chosen to have
properties similar to the Milky Way dwarf spheroidals
\citep{mateo_araa}, with  a core radius
of 0.1 kpc and a tidal radius of 2.0 kpc. (The chosen core radius is
at the small end of the range of the dSphs to compensate partially for
the lack of self-gravity in our simulations. See the discussion below
of the potential gradient across the satellite.) The projected central
velocity dispersion of the satellite is 8 km/s and the velocity dispersion
of all particles in the satellite is 6.5 km/s.
 
With 200,000 particles per satellite and a model composed of 100
destroyed satellites, the density of particles in the model
approximately matches the density of halo turnoff stars seen in our
survey. This near one-to-one relationship between particles and stars
is important because the velocity signature of kinematic substructure
may be due to the presence of only a few satellite stars in a field.
If these stars occur at high velocities (as might be expected for a
plunging orbit crossing our survey volume) then their presence in the
wings of the velocity distribution leads to a statistically significant
detection. Under-sampling the number of tracers in our simulations
would introduce biases in the detection probabilities.
 
Rather than use an N-body method to model the interaction of satellite
particles, we neglect the self-gravity of the satellite, and 
assume that each satellite is disrupted at its first pericenter
passage. Thus each simulation was started with the satellite at
pericenter.  While it is not physical that all the satellites modeled
would have become unbound on their first perigalactic passage, it is a
reasonable simplifying approximation to make for this study. Our aim
is not a detailed investigation of the tidal disruption of satellites and the
resultant tidal streams \citep[eg][]{jhb96,helmi99}, or to make a
detailed model such as has been done for the Sagittarius dwarf
\citep{jsh95,helmi2000,jib2000}.  Our aim is rather to study the
observability of tidal streams in a statistical sense.
 
In reality, satellite destruction depends strongly on the initial
conditions, particularly the structure of the dwarf galaxy and the
distribution of stars, gas, and dark matter within it.  Unfortunately
we have little information on primordial dwarf galaxies. The
properties and orbits of existing dwarfs around the Milky Way may be
special in allowing them to survive, thus telling us little about the
initial properties of the destroyed satellites. For gas-rich
satellites the loss of their gas during disk plane crossing could lead
to much of their stellar mass becoming unbound.  Our simulations are
closer to this case than to pure tidal stripping where stars are lost
over a number of perigalactic passages. In the tidal stripping case
the gradient of the potential across the tidal diameter dominates the
energy distribution of the unbound stars and the subsequent evolution
of the tidal stream.  This is because stars at the tidal boundary have
close to zero velocity relative to the satellite's center of mass.
\citep{tre93,kvj98}.  In our models the disruptive effect of the
Galaxy's tidal force is aided by the satellite's lack of self-gravity.
Thus the energy spread of the particles should be somewhat larger
than in the purely tidal case where the spatial dispersion of the
satellite correspondingly more rapid.

Streams of tidal debris were created by evolving a satellite whose
center of mass is initially along each of the 180 orbits. The
evolution of the particles in each satellite was followed for $10^{10}$
years, and the results were saved every $5 \times 10^{8}$ years. The
resulting library contains 3600 snapshots of satellite destruction
which can be sampled to create halos.  A 7th order Runge-Kutta
integrator was used for the orbit calculations.  Typical energy
conservation per particle was 0.01\% or better over the 10 Gyr.

\subsection{Comparison with N-body models}



If we are to have confidence in our method, it is important that we
understand how the neglect of the self-gravity of each satellite affects the
observed spatial and kinematic distribution of particles.  The orbital binding
energy is $U_{orb} = M_{sat} \Phi_{gal}$, while the self-gravitating binding
energy $U_{bind}$, for the spherical Plummer model, is given by

\[
\begin{array}{ll}
U_{bind} & = \, \, \frac{1}{2} \int^{\infty}_{0} \, \rho_{sat}(r) \, \Phi_{sat}(r) \, 4\pi r^2 dr
\\
\\
  & = \, \, \frac{-3\pi}{32} \, \frac{GM_{sat}^2}{b}
\end{array}
\]

\noindent For the typical system in this study, $U_{bind}/U_{orb} \sim 1/100$, suggesting
that the evolution of a bound satellite will be dominated by tidal effects.

The energy of a star which escapes the satellite is $\pm \delta E$ from the
orbital energy $E_{orb}$ of the satellite.  For a bound satellite whose mass is
much smaller than that of the host system, $|\delta E/E_{orb}| \ll 1$
\citep{jhb96,kvj98}.  Neglecting self-gravity is equivalent
to scaling up $|\delta E|$, which tends to disperse the satellite
more rapidly into the available phase space.  Therefore the validity of this
approximation hinges upon the relative debris dispersal in the two cases.

To test the approximation we ran four representative N-body
simulations for 10 Gyr using the tree code of Hernquist (1987; 1990),
with $N$=1000 and $M=10^7$ M$_{\odot}$. In this case each body
represents $10^4$ stars. While we expect that some features of the
tidal debris will be dependent on $N$, the disruption of the satellite
depends foremost upon $U_{bind}/U_{orb}$, which we preserve by
construction.  Each of the simulations preserved the total energy to
$|\Delta E/E| < 10^{-3}$.  The four simulations we performed spanned
the range of (absolute) $E$ and angular momentum $J_{tot}$ for the
ensemble: high $E$ and low $J$ (simulation 1039), high $E$ and high
$J$ (1213), low $E$ and low $J$ (1250), and low $E$ and high $J$
(3117).  The final spatial and kinematic configurations of the
self-gravity and non self-gravity treatments are shown below in
Figures \ref{jsax} and \ref{jsav}, respectively.  To facilitate the
comparison we have plotted a random subset of 1000 of the $2\times
10^5$ stars in each non-self-gravitating case.  Figure \ref{jsax}
shows the projection onto the meridional plane and the z-projection
for each model.  Each plot of the meridional plane includes the
zero-velocity surface for the dynamical center of the satellite system
to aid in the comparison.  Figure \ref{jsav} shows the $V_z$ and $V_y$
projections for each model.  The distributions resulting from the two
treatments of self-gravity are essentially indistinguishable for the
low $J$ models (1039 and 1250).  The morphology of the debris in
simulation 3117 is similar in the two treatments, although the
particle density along the tidal stream is lower in the self-gravity
case.  The self-gravitating model 1213 populates a much smaller volume
of the phase space available compared to its non-self-gravitating
counterpart.

The ability of a satellite to fill the available phase space depends upon the
rate at which the satellite disrupts.  Figure \ref{jsabound} shows the time
evolution of the bound fraction of each satellite.  The low $J$ simulations
1039 and 1250 pass within $\sim 3$ kpc of the Galactic center, where the
tidal field is strongest, and disrupt very rapidly, thereby spending most of
the duration evolving as would the non-self-gravitating models.  Model 3117
sheds about 40\% of its mass in 10 Gyr, resulting in a tidal stream density that
is correspondingly lower than the non-self-gravitating model (except
$x,y,z=-15,2,-21$ kpc, where the bound particles and the most recent escapers
still aggregate).  Model 1213 still contains almost 3/4 of its original
membership after 10 Gyr, and so differs most from the initial total dispersal
approximation of the non-self-gravity model.  Note that the outer fifth of each
system begins the simulation unbound because the tidal radius used to construct
each truncated Plummer system is set to 2 kpc, regardless of the orbital
parameters.


The validity of our approximation depends upon the particular dwarf galaxy
orbit.  It is clear that the approximation is valid for low $J$ orbits.  In the
case of high $J$ orbits, we have a problem in that the models 
may overestimate the density of stars
along the tidal debris stream, especially for high $J,E$ orbits.  However,
since the sampling density of orbital parameters for the dwarf ensemble is
lowest toward high $E$ and high $J$ systems (see the bottom panel of Figure 1), the
problem is reduced somewhat.  In addition, the presence of moderate amounts of
gas in each primordial dwarf can significantly shorten its disruption time,
since gravitating gas would be stripped in short order, further mitigating the
problem.  In any case, our ignorance of the initial state of each dwarf renders
the uncertainty in satellite disruption times of a small subset of these dwarfs
a second-order problem to which we will return in a subsequent study.

\section{Views of the models}

\subsection{Views from a theoretical perspective}



The evolution of a satellite on three orbits with large apocenters is
shown in Figure \ref{xyzgalclean}. The left two panels of each row
show the XY and ZY spatial projections of each orbit. The heavy black
locus of points shows the satellite after 1 Gyr of evolution
subsequent to its first perigalactic passage. The lighter points show
the results after 10 Gyr.  The right hand panel shows the radial
component of velocity and distance with respect to the Galactic
center.  The three different simulations are chosen to have an orbital
pericenters between 8 (top) and 2 (bottom) kpc. The latter reaches
regions where the potential gradient is greater. The orbital period is
smaller for the bottom simulation, leading to more frequent
perigalactic passages. This orbit also spends more time closer to the
influence of the disk. All these factors contribute to the increased
dispersal of the satellite.

Figure \ref{xyzgalmessy} shows three satellites on orbits with smaller
apocenters at the same two stages of evolution as Figure
\ref{xyzgalclean}. For these orbits, spatial structure is rapidly
dispersed and little remains after only a few Gyr.  However, the
velocity structure remains visible in the right-hand panels of both
Figures. While it is possible to directly associate each wrap of the
orbit seen in the spatial projections on the left with the loops in
phase space on the right of Figure \ref{xyzgalclean}, it is seen that
this is not possible in Figure \ref{xyzgalmessy} as most of the
spatial information has been lost.

The power of searches in velocity space to detect older accretion
events is clearly seen.  The reason for the persistence of structure in the
velocity vs. Galactocentric distance diagram is that we are plotting
conserved or nearly conserved quantities. The apocenter of the orbit
reflects the particle's total energy, and pericenter is dominated by
the particle's total angular momentum. While only the z component of
angular momentum is strictly conserved for an axisymmetric potential,
the dispersion in total angular momentum of satellite particles is
relatively small if their orbits remain outside the disk region. The
spread in total angular momentum for the particles in Figure
\ref{xyzgalclean} is approximately 20\% larger than their initial
spread. This would be larger if the halo potential was significantly
flattened.

\subsection{Views from an observational perspective}

Our perspective as observers is limited to the view of the Galaxy from
the Sun. 
In theory, if distances were known accurately and proper motions were
available, we could transform back to the Galactocentric perspective.
However, distances based on broad band photometry of halo turnoff stars
are at best accurate to 50\% due to their near vertical evolution in
the color magnitude diagram.  These stars are also too faint and
distant to have currently measurable proper motions of useful
accuracy\footnote{ a
  transverse velocity of 150km/s, typical for a halo star, gives a
  proper motion of 2mas per year for a star at a distance of 15kpc,
  which is currently only measurable, even from space by Hipparcos,
  for much brighter stars  --- we will
have to wait for the next generation of astrometric satellites to fill
in the missing information.  In the meantime we can use models of
satellite destruction transformed to the solar perspective to help
in detecting and understanding observations of substructure.}. In this
paper we will thus concentrate on radial velocity measurements alone.



Figures \ref{lbsunclean} and \ref{lbsunmessy} show the same
satellite orbits as Figures \ref{xyzgalclean} and \ref{xyzgalmessy}, 
but plotted against observable quantities to illustrate
the effects of projection on the appearance of the orbits. (At this
stage, we do not add the effect of observational error on distance and velocity.)

The left panel shows the distribution of the disrupting satellites
over the sky in $l$ and $b$.  The center panel shows the relation between
$R_{\odot}$  and longitude, and the right panel the
heliocentric radial velocities versus distance from the Sun.  Even the
three relatively simple  debris streams seen in Figure
\ref{xyzgalclean} become more difficult to interpret with the shift in
perspective from galactocentric to heliocentric.  

This is further complicated by the distance limit of any velocity
survey, optimistically set here at 30 kpc which corresponds to
approximately V=21.5 for halo turnoff stars.  For orbits with large
mean radii, only the portion of the orbit near pericenter is visible
causing the streams of tidal debris to appear as isolated islands. The
density of particles is also reduced due to the relatively small
fraction of the orbital period that particles spend near pericenter.
However, the velocity substructure signal remains clear in the right
hand panel of Figure \ref{lbsunclean}, especially for those particles
more than approximately 10 kpc from the Sun.

For the orbits with smaller mean radii, we have the advantage that
most of the orbit remains within the survey volume, but this is offset
by the more rapid mixing of the satellite particles.  Orbit 1082, seen
in the lower panel of Figure \ref{lbsunmessy}, represents a
particularly extreme case where no velocity substructure information
apparently remains to isolate the satellite debris.  The other two
orbits in Figure \ref{lbsunmessy} show more promise for detection via
velocities. The individual wraps of the orbit are no longer clearly
separated as was seen in the right hand panel of Figure
\ref{xyzgalclean}. This is because projection effects due to our
observing perspective from the Sun are larger than the separation
between wraps. However, the satellite particles still occupy a
relatively narrow region of phase space.

\section{Pencil-Beam Surveys}

As we shall see below, observations in discrete
fields, such as those from pencil beam surveys, can recover more of the
information contained in phase space than appears likely from the
right hand panel  of Figure \ref{lbsunmessy}. 
In this panel, two of the three spatial coordinates
have been summed over, obscuring the correlations of radial
velocity with spatial structure seen in the other two panels.  This
holds true even for orbits such as 1082 after it has mixed for 10 Gyrs.
We will show below that the observed velocity distributions, even in
such well-mixed cases, can vary significantly from those
expected from a smooth halo.  

Now we consider observations in individual fields.
Figure \ref{rvlb1messy} shows the spatial and velocity distribution of
the particles from satellite 1082 seen at age 1 Gyr.  The bulk of the
satellite particles in this figure  are distributed
between two consecutive apocenter passages of their orbit. In order of
decreasing energy (and increasing $l$), particles are distributed from
an apocenter of 28 kpc near $l=310$, through pericenter at $l=10$ to
an apocenter 19 kpc at $l=150$. The rest of the particles with higher
and lower energy spread over an additional two wraps of the orbit but are
few in number as they originate in the outer, low density regions of the
satellite.

In the lower section of Figure \ref{rvlb1messy} the velocity
histograms are shown for three lines of sight. Most of the
observations of this satellite will show a larger velocity spread than
in the original satellite. This is because most lines of sight will
intersect a significant fraction of the orbit since its apocenter
is small. The two histograms on the right are broadened due to
observing particles near apocenter, both coming and going. The
histogram on the left shows one of the few sections of the orbit where
a narrow velocity dispersion is seen.  However, this case, a low
energy satellite observed at only 1 Gyr after first passage, is less
relevant to our question of late halo building because satellites with
such small mean radii were most likely accreted very early in the
Galaxy's history, before most of its mass was acquired.



When the majority of the satellite has not spread much beyond a single
orbital wrap it is relatively easy to trace the relationship of the
features seen on the sky to the velocities, for example, in Figures
\ref{rvlb1messy} and \ref{rvlb1clean}, for satellites 1082 and 1197
respectively.  
However, when the debris have wrapped many times around the orbit (as
will be the case for debris that spends most of its time near the
solar circle) this is no longer possible.  Satellite 1082 is shown in
Figure \ref{rvlb2messy} after 10 Gyr of evolution. The satellite
debris still traces a fairly narrow path across the sky but at each
position there is a large range of velocities (and distances)
present.  
Despite the fact that the particles have wrapped 13 times
around the orbit, the velocity histograms here are distinctly
non-Gaussian. Velocity structure remains, despite the very smooth
spatial appearance. In reality, small effects such as scattering off spiral
structure and molecular clouds in the disk will add further to the
smearing in velocity space.

We are using these satellites as examples because their confinement to
a narrow band of $b$ on the sky makes it easier to produce an
understandable two-dimensional figure. However, it is somewhat
misleading, in that only for orbits in the plane of the disk are the
debris confined to such a small volume. In general the orbital plane
will precess and thus the debris will spread over many orbital wraps
and will eventually fill a torus-like volume (as can be seen in Figure
4 of \citet{helmi99}). It is then much less likely that a narrow line
of sight will intersect multiple wraps of the same satellite and hence
the observed velocity distributions will be narrower and mono-modal. A
velocity-distance plot, made possible with high accuracy proper
motions and parallaxes from Hipparcos \citep{helmihip}, is then more
illuminating.



In Figure \ref{rvlb1clean}, which shows satellite 1197 with
apocenter 100 kpc, it can be seen that once a tidal feature is
detected in a single field, it will be possible to trace it on the sky
in other fields, as there will be a clear correlation of velocities
with position on the sky (and distance, as can be seen in Figures
\ref{lbsunclean} and \ref{lbsunmessy}).  Even with limited distance
information it should be possible to constrain the orbit
of the satellite with only the radial component of the velocities
measured.  Figure \ref{rvlb2clean} shows the satellite particles after
10 Gyr of evolution. The more complicated spatial and velocity
structure is due to the particles having wrapped five times around the
orbit --- most of the particles on each wrap are more than 30 kpc from
the Sun and are thus not plotted. Along many lines of sight, particles
on different wraps of the orbit are seen with significantly different
velocities.  However, it will be possible but more difficult to trace
it on the sky by following the run of velocities with position and
distance. 

In summary, despite difficulties introduced by the Sun's position
relative to the stream and our survey distance limits, velocity
structure remains. Pencil-beam surveys can detect this structure even
if the spatial density of the stream is significantly reduced by its
evolution in phase space, and the situation is further confused by the
appearance of multiple wraps of the stream along the line of sight.

\section{Observing single strands against a smooth halo.}

We have been considering the properties of tidal debris from
individual satellites in isolation. However, we know that in the solar
neighborhood there is a well-mixed component of the halo \citep[as can
be seen in][]{helmihip} which will complicate our detection of
satellite debris. It is possible that the outer halo, where few field
stars are known, is dominated by tidal debris, but this part of the
Galaxy needs to be surveyed using the more luminous but rarer red
giants and so a different detection strategy will be needed.

We will, thus, consider the case of detecting the debris from a single
satellite seen against a smooth well-mixed halo.  The advantage of
this approach is that we can represent the kinematics of the
well-mixed component using a velocity ellipsoid, rather than having to
evolve the orbits for each particle. The observed distribution of
radial velocities along a given line of sight is close to Gaussian
even for a non-isotropic velocity ellipsoid as long as the halo does
not have significant net rotation \citep{cl86,jen86,beers2000}. Thus
the null hypothesis which we test against (Gaussian shape) is
well-defined, and there already exist efficient statistical tests.
We have assumed that the contributions of the thin and thick disk 
populations to the velocity distributions can be ignored.
This is true for our survey, where the photometric accuracy of the
photometry combined with the color and magnitude range used to define
the main sequence turnoff region minimizes any contamination by disk
stars \cite{spag1,spag2,spag4}. However in the case for studies
based on photographic photometry \citep[eg][]{gw95} 
the larger photometric errors will lead to significant disk
contamination.

We populate the smooth halo along each line of sight according to an
$r^{-3}$ density distribution.  The normalization is set to match the
density of halo turnoff stars seen in the solar neighborhood
\citep{bc86,spag1}. Although there is increasing evidence for a
moderate flattening in the inner halo \citep{tdk65,psb91}, 
we have chosen for simplicity
to use a spherical model. The spherical halo model over-estimates the
number of stars at the pole, and underestimates the numbers towards
the anticenter compared to the observed flattened halo. A moderately
flattened halo would not produce significantly different answers in
what follows. The velocity ellipsoid used, from which the observed
radial velocities of the smooth halo particles are sampled has
$(\sigma_{r}, \sigma_{\phi}, \sigma_{\theta}) = (161,115,108) $
\citep{cy98}.  We also use an isotropic velocity ellipsoid with
$(\sigma_{r}, \sigma_{\phi}, \sigma_{\theta}) = (115,115,115) $;
closer to the values measured in more distant halo fields. Both models
have a mean rotational velocity of zero.  (The exact underlying
distribution is known in our models, but there are currently few
observational constraints on the true distribution of halo velocities
away from the solar neighborhood.  Even the mean rotational velocity
is subject to some dispute \citep{srm92}.)

We use a grid of 61 fields with $ 30 < b < 80 $ and $ 0 < l < 180 $.
Their distribution is shown in Figure \ref{fieldlb}. We need consider
only one quadrant of the sky because of the inherent symmetries of the
system, discussed below.  The debris from each satellite is observed
at each of the 20 snapshots spaced 0.5 Gyr apart that cover the
evolution from 0.5 to 10 Gyr.  Due to limitations on current
facilities for spectroscopic followup of faint stars, our optimistic
survey distance limit used above of 30 kpc has been reduced to 20 kpc.
The 180 initial satellite orbits all have pericenters less than 27
kpc.  Debris from satellites on orbits with pericenters larger than
this is almost never detected.


We are interested in what makes an orbit detectable, not the accidents
of viewing geometry. Therefore we will average the detection
probabilities of each strand over 100 realizations of observations.
These randomizations also minimize any sampling biases caused by a
fixed grid of fields and increase the volume of phase space occupied
by the orbits. The following parameters are varied for each
observation:
\renewcommand{\labelenumi}{(\roman{enumi})}
\begin{enumerate}
\item Viewing orientation of the strand is altered by selecting at
  random the azimuthal position of the Sun in the X-Y plane through
  $0-360 ^\circ$.  There is nothing special about the azimuthal position
  of the Sun with respect to a satellite's orbit. The relative
  orientation of Sun and particle orbit alters which particles are
  included in the survey volume and how the components of their space
  velocity are projected into their observed radial velocity.
  
\item The galactocentric radius of the Sun is varied randomly from 8-9
  kpc. This modifies the line of sight through debris, particularly
  those particles that pass close to the Sun.
  
\item The actual field center used is offset at random by one degree
  on the sky. This alters the line of sight through more distant
  orbits and minimizes the possibility of chance alignments of the
  limited sample of orbits with the grid of field positions, which
  would bias our results.
  
\item The orbits of the particles are reflected randomly about the
  Galactic plane. This ensures our fields are representive of both
  Galactic hemispheres.
\item The orbital direction of the particles are reversed randomly.
  This ensures velocity symmetry.
\end{enumerate}

In most cases few, if any, particles from a single strand are present
in a given field due to the small filling factor of the debris.  The
variation in the distribution over the sky of the debris was seen in
Figures \ref{rvlb2messy} and \ref{rvlb2clean} where two strands with
orbital periods of  0.26 and 1.1 Gyr respectively are shown. It can be
seen that the shorter period orbit satellite particles are almost
completely disrupted and have a relatively large filling factor,
occupying most of the volume that its precession traces out.

If five or more particles are present in a field, we test for their
detection\footnote{Only running the test when five or more particles
  are present significantly reduces the computing time required with
  little loss of accuracy --- less than one percent of detections
  occur when there are less than five particles present from the
  strand.}. Because we do not want results dominated by random
effects, we construct 25 realizations of a smooth halo in that
direction and add to each the strand particles from the field.  A 20
km/s Gaussian velocity error is also added to the particle velocities to match
our observational errors.

\subsection{Statistical Tests for Substructure}

The velocity distributions of the smooth component of the halo appear
close to Gaussian for both the isotropic and radially elongated
velocity ellipsoids. In the latter case the line of sight projection
of the three components of the velocity ellipsoid varies with distance
--- the radial component becoming dominant at large distances
\citep[see][for a derivation]{woolley}.  The standard deviations range
from 120 to 160 km/s depending on the Galactic coordinates of the
field.

We will show below that, contrary to naive expectations that the
satellite debris will produce a single narrow velocity peak on a
smooth distribution, there are many different signatures of
substructure, including multiple peaks and broad but asymmetrical
velocity distributions. The statistical test we use must accommodate a
large range of deviations from Gaussian shape, and should not be
focussed too narrowly on a particular velocity signature. 

We use Shapiro and Wilk's $W$-statistic to test for substructure in
the combined velocity histograms because of its omnibus behavior ---
it is sensitive to many different deviations from Gaussian shape. The
$W$-statistic is a sensitive and well-established test for departures
from Gaussian shape \citep{sw65}, which is based on a useful technique
of exploratory data analysis, the normal probability plot. This plot
transforms the cumulative distribution function (CDF) of a gaussian to
a straight line so that deviations from gaussian shape are immediately
noticeable.  The $W$-statistic can be viewed as the square of the
correlation coefficient of the points in the probability plot: the
closer the data approach a straight line on the probability plot
the closer the $W$-statistic is to 1. The confidence level that the
distribution is non-Gaussian ($P$-value) can then be calculated based
on the $W$-statistic. \citet{roy95}
\footnote{The fortran subroutine from the journal article can be downloaded
from http://lib.stat.cmu.edu/apstat/R94 } provides an
algorithm for evaluating the $W$-test, and estimating its $P$-value
for any $n$ in the range $3 \le n \le 5000$. 

\citet{the book} give a critical summary of goodness of fit tests and
comment on the efficiency of the Shapiro-Wilk $W$-test: the test is a good
compromise between tests requiring too many assumptions about the
distribution and overly general tests which have little power.
Examples of the former include tests based on detailed knowledge of
the moments of the distribution, and the latter, the Kolmogorov-Smirnov
test. 

The $W$-statistic responds to the minor deviations from Gaussian shape
in the smooth distribution caused by the non-isotropic velocity
ellipsoid. This results in a small bias in the $P$-values which varies
little from field to field. At the 99\% confidence level there is a
failure rate of 0--1\% more than the 1\% that would be expected for a
pure Gaussian distribution.  This bias is small compared to the effect
of the small filling factor of the debris and will be left
uncorrected.  Smooth halo models were also run with an isotropic
velocity ellipsoid and there was little change in the detection
probabilities (see below).
 
In our analysis of the simulated velocity distributions we define a
detection to be a rejection of Gaussian distribution of velocities at
the 99\% confidence level ($P$-value $< 0.01$), when there are five or
more satellite particles present in the field.

\subsection{Detections of tidal debris against the underlying halo}

Before compiling the overall detection probability of satellite
debris, it is worthwhile to develop an empirical understanding of the
range of velocity distributions that can lead to a detection. Earlier
sections looked only at the behavior of satellite debris in isolation,
without the presence of the underlying smooth halo particles and
observational errors; also there were no selection criteria which
correspond to real observational situations such as pencil-beam
surveys.

Examples of velocity histograms that pass our detection criteria are
shown in the lower half of Figure \ref{hist32}. Each of the four
columns show data from a satellite detected in different fields on the
sky. The unshaded histogram gives the combined distribution of
satellite and smooth halo particle velocities observed in the field.
The shaded histogram shows the velocities of the satellite particles.
The (l,b) coordinates of the field are shown on the upper left of each
velocity histogram and the $P$-value of the detection on the upper
right.  The upper half of Figure \ref{hist32} shows the magnitude
distributions of the same particles. The (l,b) coordinates of the
field are again given at top left, and the age of the satellite debris
in Gyr at top right.  The velocities include an observational error of
20 km/s and the observed magnitudes include a sigma of 0.35 mag,
corresponding to the scatter in absolute magnitude of halo turnoff
stars within the color range of our survey \citep{spag1}.

The velocities of the satellite particles show a broad range of
properties. Orbit 1197 is the simplest to interpret --- the four
histograms each show a single peak in velocity far away from the mean
of the smooth halo for the satellite
particles due to a single intersection of the orbit of debris and the
line of sight. In this case the detection probability depends strongly
on the number of stars in the wings of the velocity distribution.
The detection at (135,30) is marginal because there are only 6
satellite stars in the field. 
With a mean orbital radius of 63 kpc, the debris are
outside the survey volume 90\% of the time, minimizing the chance of
multiple wraps being seen in a field. This panel is typical of the
detections of other satellites on orbits with large mean radii.

Orbit 6866 (180,80) shows a different detection situation: the satellite
particles have a velocity close to the mean but the large number of
satellite particles in the peak leads to a symmetric but very
non-Gaussian velocity distribution. Thus the hypothesis of Gaussian
shape is rejected with very high confidence.

Column 4 of Figure \ref{hist32} shows 
another unusual case: a polar orbit similar to that proposed for
the Sgr dwarf \citep{helmi2000}. Because of its polar orbit it has
remained confined to a plane and the probability of detection of
satellite particles on multiple wraps is proportionately higher.

The histograms of the right-hand three orbits of Figure \ref{hist32}
 show multiple peaks in the
velocity distribution of satellite particles.  This is due to the line
of sight intersecting multiple wraps of the debris.  Each peak in the
velocities of orbits 6866 and 1031 are from particles on a single
wrap. However, interpretation is more complex for orbit 1082:
particles on each wrap of the orbit contribute to each of the peaks.
This is due to the apocenter of the orbit (22 kpc) falling within the
survey volume.

Figure \ref{erm} illustrates this situation by showing the
relationship of total energy to Galactocentric radial velocity: energy
changes slowly and almost monotonically along a single wrap.  The
sorting in energy of the particles due to phase space conservation is
clearly visible.  The crosses show where particles seen within the
field originate in overall energy distribution of the debris
particles.  In orbit 1082, the particles come predominantly from two
wraps, but because the apocenter of this orbit is only 22 kpc, we see
particles over most of the orbit in a single field (both ``coming''
and ``going''). In order to have enough particles from this diffuse
stream to trigger a detection in the velocity histogram, orientations
where streams are lined up close to the line of sight are needed.

The debris on orbit 1082 mixes rapidly, leading to
approximately 13 wraps of the debris after 10 Gyr.  The velocities of
satellite particles tend to be closer to the underlying distribution
of halo velocities and hence more particles are usually required for a
detection.  The detection shown in Figure \ref{hist32} at 9 Gyr for
this satellite is close to the limiting detection threshold of 1\%
despite the presence of 50 satellite particles.



After we have detected substructure, a natural next step is to attempt
to determine the orbits of the particles and then to reconstruct the
properties of the satellite from the debris \citep[eg][]{helmi2000}.
In extreme cases it is clear which particles in the combined
distribution belong to the satellite (for example orbit 1197).
However, this is not the case for most of the detections. The combined
velocity distributions are often clearly non-Gaussian to the eye, but
it is not easy to accurately identify the contribution of the
satellite particles from either the velocity or magnitude
distributions alone. While in a single field it is
not possible to uniquely identify which stars are from an accreted
satellite, it is possible to make statistical allocations. 
We are exploring the use of mixture modeling on
the velocity distributions of each field to quantify the components
present in the distribution \citep{smhw}.

The fraction of the halo that has been accreted can be estimated using
the fraction of fields that show kinematic substructure.  This can be
modeled by extending our simulations of model halos to those
composed of varying mixes of smooth component and debris from multiple
satellites. The accreted fraction can be estimated more directly, once
the kinematic and spatial distribution of stars in each stream is sufficiently
well quantified to allow an estimate of the properties of the
progenitors \citep[eg][]{helmi2000}.  

It is tempting to think that observations made with higher velocity
accuracy would simplify the task of identifying the satellite
particles. This would be true if our line of sight was nearly normal
to a stream that is well phase-mixed \citep{helmi2000} in which case
the velocity variation along the stream would only be detectable with
proper motions.  However, that situation is geometrically rare, and the
number of observable stars from the satellite within a field will
usually be too small to trigger a detection based on the velocities.
In fact, there is a bias towards detecting streams of debris that
obliquely cross the line of sight due to the higher projected density
of satellite particles. In this case the width of the observed
velocity distribution from a single wrap will usually be
dominated by the velocity variation along the stream. This can be
clearly seen in the middle right panel of Figure \ref{erm}.

Combining velocity data with even inaccurate distance information is
helpful.  This is seen in the upper panel of Figure \ref{erm} for both
satellites. The clear correlation of velocity with distance seen in
the middle panel of Figure \ref{erm} for both satellites remains when
the distances are transformed to magnitudes (including the absolute
magnitude distribution) and 20 km/s errors added to the
velocities. The linear dependence of velocity on distance can still be
seen in the upper panel. 
(Deriving distances from observed magnitudes and
colors leads to $\sim$50\% distance errors.)

\subsection{Detection probabilities}

We are now in a position to combine the above work into quantitative
estimates of detection probabilities of debris.  Our aim is to
determine the probability of detecting velocity substructure from the
debris from a single satellite on a given initial orbit when we
observe velocities for a sample of halo turnoff stars in a single
field.  We consider the example of a single night's observing with a
multi-object fiber spectrograph on a single field of a preselected
sample of halo turnoff stars (uncontaminated by thin or thick disk
stars). The detection probabilities derived below are representative
of the likelihood that these observations would show substructure in
the velocities. Such an observation would typically return from 50 to
several hundred halo turnoff star velocities, depending on the
telescope/instrument combination.

There are two major contributions to the detection probability: first
there need to be satellite particles in the field observed, and the
probability of this is quite small. Second, if particles do exist in
the field, we calculate the probability that they are detected in a
velocity histogram. 

In the following subsections we will look at how the detection
probabilities vary with the properties of the telescope and
instrumentation used for the velocity measurements.  In particular we
consider the influence of the number of stars observed per field, the
limiting magnitude, velocity accuracy and location of the fields
observed on the substructure detection probabilities.

\subsubsection{Velocities obtained for all halo turnoff stars}


With instruments such as the Anglo-Australian telescope's 2df system
or the Sloan fiber spectrograph, there are sufficient fibers to
observe almost all halo turnoff stars in a field of size one square
degree.  (Numbers of turnoff stars per square degree brighter than
V=20 will vary from 100 to 500 depending on Galactic latitude and
longitude \citep{spag1}.)  In this case the typical detection
probability of a single strand 
is $\simeq 1\%$.  Figure \ref{pcondc}
summarizes the contributions to detection probability for each of the
180 satellite orbits in the library if they were to be observed 6.5 Gyr
after first pericenter passage. The number of times that five or more
particles are ``found'' in the field is dominated by the fraction of
the orbit accessible within the survey volume, as seen by a comparison
of the upper two
panels.  The small filling factor on the sky, particularly of orbits
with large mean radius, causes this percentage to be low.  The third
panel shows the probability, given that five or more stars are in the
field, that the strand is detected in the velocity histogram.  The
bottom panel shows the total detection probability plotted against the
mean radius of the satellite orbit. This is obtained by multiplying
the probability that satellite particles are present by the
probability that if particles are present, they are detected. Thus the
detection probability in the bottom panel is the product of the
probabilities in the two panels above.

A general decrease in detection probability for mean radii above 15
kpc is seen. This is primarily due to the small fraction of time such
satellite debris spend in the survey volume.  Once particles from the
satellite are found in the field, the detection probability is
relatively high for all orbits except the innermost ones.
The particles with orbits of large mean radius are relatively close to
pericenter when detected, and thus their large radial velocities are
easier to detect against the smooth halo velocity distribution, even
when few particles are present, as can be seen in Figure \ref{hist32}.
The detection probability drops rapidly for orbits with mean radii
below 15 kpc for three reasons. These orbits have shorter periods and
thus disperse more rapidly. Thus although particles on such orbits
pass through the survey fields frequently, $\sim$30\% of the time, the
detection probability is low since there are usually too few particles
present in a field for a significant detection. Also their velocities
are on average less extreme and so harder to detect against the
underlying distribution of velocities.  Orbits with mean radii below
10 kpc have dispersed sufficiently that the probability of five or
more particles occurring in a field is low.

The orbits that are confined close the plane have low detection
probabilities as we have no fields below 30 degrees. There was no
pre-selection made against these orbits in our simulations since in
reality it becomes increasingly difficult to detect halo stars
reliably in fields with high stellar density and variable reddening.

\subsubsection{Time dependence of detection probabilities}


We now explore the dependence of detection probability on time since
satellite dispersion.
Figure \ref{pagead} shows the detection probabilities at ages 1.5,
4, 6.5 and 9 Gyr. In order to show the general trends the detection
probability for each satellite has been averaged over a 2.5 Gyr
period.  It is striking how little the detection probabilities vary
with time.  As the strands evolve, the mean densities along the
strands decrease by factors of 100 to 1000.  The competing effects of
spatial spreading, which increases the probability of finding particles from
a strand in a field, and the decreasing mean number of particles found
in a field approximately balance.  The general trend is that more
tightly bound orbits become more difficult to detect with time, due to
their more rapid spatial dispersion decreasing the number of stars per
field below a level that triggers a detection.  Conversely, the orbits
with large mean radii become easier to detect at later times.


This behavior is seen in Figure \ref{pfd_cy20} where the ``found''
and detection probabilities are shown at each time step for the debris
from six satellites.  For example orbit 6866 shows the trend clearly
with a steady increase in the ``found'' probability reaching 35\% at
10 Gyr as the debris disperse spatially. However, the fraction of the
``founds'' that result in a detection decreases from 55\% at 1 Gyr to
8\% at 10 Gyr. The broad peak in the detection probabilities is the
result of these two effects.  As the mean radii of the orbits
increase, the peak in the detection probabilities moves to later times
and broadens. Orbits with mean radii beyond 35 kpc have detection
probabilities that on average remain constant.  The main source of
variation in their detection probabilities seen in the upper three
panels of Figure \ref{pfd_cy20} is caused by the presence of
densest parts of the tidal debris  in the sample volume.

\subsubsection{Detection probabilities with smaller samples of halo turnoff stars}

For systems like the NOAO Hydra spectrographs with 98 or 132 fibers it
is only possible to obtain velocities for 50-100 halo turnoff stars in
a single field with an exposure time of 6-8 hours. Also, unless
observing conditions are perfect it is impossible to obtain accurate
velocities for the more distant candidates in the sample.

To test the effect of these observational constraints the velocities
in each field were sub-sampled by randomly selecting the required
number of velocities from the combined sample of satellite and smooth
halo velocities. A new subsample was created for each of 25
realizations of the smooth halo. As before this was only done when 5
or more satellite particles were present in the full data set for the
field.  An added constraint was that four or more satellite stars had
to remain in the velocity subsample before the $W$-test was run
(because the number of satellite stars varied in each subsample).

A comparison of the detection probabilities averaged over all ages for
the cases where velocities of all\footnote{number of halo stars per
field will vary from $\sim$100--500, see section 4.2.1} halo stars detected in the field, or
100, or 50 halo stars are obtained within each
field is shown in Figure \ref{prob3_cy20}.  It can be seen that the
overall shape of the relationship between detection probability and
mean radius of the orbit remains basically unchanged, but the
detection probabilities scale roughly with the number of velocities
obtained. This general behavior is true at all of the ages sampled.
Both the number of fields with satellite particles found and the
probability of detection if such particles exist scale down in
approximately the same way.


In summary the detection probabilities show a linear decrease with the
number of stars observed per field, falling to $\sim 0.25\% $ (for debris
on orbits with mean radius above 30 kpc) when only 50 halo turnoff
stars. It is important to obtain velocities of as many halo
turnoff stars as possible within the field to maximize the detection
probability. 

\subsubsection{Detection probabilities at brighter limiting magnitude}

Another series of observations of the models were made with the survey
radius reduced to 12 kpc, corresponding to a magnitude limit of
approximately V $=$ 19. This represents the current magnitude limit
achievable with Hydra on the WIYN telescope in 6 hours of exposure
for 20 km/s velocity accuracy on halo turnoff stars with M$_V \simeq 4$

Figure \ref{prob3_cy12} shows the detection probabilities averaged
over all ages for the three cases where all, 100, and 50 velocities are
obtained within each field. These probabilities should be compared to
those in Figure \ref{prob3_cy20}. It is seen that the upper envelope
of detection probabilities has decreased by factor of 2 -- 3,
comparable to the factor of 2.7 reduction in the volume surveyed.  The
25 orbits with pericenters greater than 17 kpc are no longer detected.


\subsubsection{Dependence of detections on $l,b$}

For the fields studied there is no strong dependence of the detection
probability with $l,b$ coordinates except for orbits with pericenters near the
survey radius. These orbits can only be detected in the fields 
towards the anticenter that probe larger Galactocentric distances.
The fields closer to the Galactic center do not sample as large
Galactocentric distances as the anticenter fields.  Thus on average,
orbits with larger pericenters are less likely to be detected in these
fields because of the distance limit of the survey.  However, they
have an increased detection probability for orbits at intermediate
Galactocentric distances. This is because it is possible to survey a
larger volume at Galactocentric distances close to the solar radius on
the opposite side of the Galactic center.

\subsubsection{Observational velocity errors}

It might at first sight seem advantageous to obtain very precise
velocities in order to isolate the stream most effectively. However,
the detection probabilities are not significantly improved for
velocity errors less than 20 km/s. The detection probabilities are
only 10\% lower on average with a 20 km/sec error than for no error.
Increasing the errors to 50 km/s is more significant and degrades the
overall detection probabilities by 30 to 100\% depending on the
distribution of satellite velocities in the field.  This behavior is
partly due to the nature of the Shapiro-Wilk test as a general test of
non-Gaussian shape, rather than as a specific test for multi-modality.
It is also due to the velocity dispersions of 20 km/sec or larger for
orbits with mean radii less than 25 kpc.  The widths are typically
caused by particles on multiple wraps contributing to a single
velocity peak.  The velocity dispersions (of individual velocity
peaks) in the detections of particles from satellite orbits with mean
radii greater than 40 kpc have a modal value of 5 km/s with a tail
extending to 25 km/s. Thus they suffer more from larger velocity
errors.

Lower errors in the initial survey velocities thus make the subsequent
identification of which stars belong to the tidal debris more
efficient.  Sharp peaks in the distribution, rather than a weak
asymmetry caused by dilution of the signal, are not only more
convincing to the eye, but are more amenable to other statistical
tests for verification.

\subsection{Detection Probabilities with multiple strands}

In cases where debris streams from multiple satellites are present in
the halo, the detection probabilities will increase with the
number of destroyed satellites present until the debris has
significant overlap on the sky and the detection rate levels out or decreases. 
If we consider just satellites with large mean radii, whose
detection probability is dominated by the fraction of the time spent
in the survey volume, then there should be a linear increase in the
detection probabilities until debris are ``found'' often enough to
overlap in the survey field. Thereafter, the occurrence of multiple satellites
in a single field will slow the increase in detection probability due
to confusion. Eventually the detection probability will start to
decrease. This becomes less likely, however, as Galactocentric radius
increases, as the decrease in overall density will balance this
confusion effect and the streams will have higher contrast.

The techniques we have used to model the detection probabilities will
need to be modified to cope with the case of a predominantly lumpy
halo, which is more likely at large Galactocentric radius.
Detection strategies modeled on searches for multimodality in
velocity space, \citep[for example][]{jiayang}, rather than the
deviation of the velocity distribution from a Gaussian shape, will be
useful.

Because of the difficulty of identifying large samples of distant halo
stars, we currently have no reliable in-situ
samples of outer halo star velocities.  There have been several
attempts to study the outer halo  via ``armchair
cartography'' \citep[eg][]{sl} --- using the properties of
outer halo stars transiting the solar neighborhood to extrapolate to
the entire outer halo. This extrapolation is risky if hierarchical pictures of
the ongoing growth of our halo by accretion of small systems are
correct, as many outer halo objects will never reach the solar neighborhood.
Large systematic surveys of outer halo objects
\citep[eg][]{spag1,yanny2000,srm00} will be needed before we can start to
address this problem.

\subsubsection{Satellites of different mass and size}

We have modeled the detection of debris from a single low-mass ($10^7
M_\odot$) satellite with an initial structure selected to match
existing Galactic dSphs at the low-mass end.  We now use our finding of
the relative independence of the detection probabilities to the time
since the satellite was disrupted, and the parameterization of the
evolution of tidal debris of \citet{kvj98} and \citet{helmi99} to
extend our results to satellites with different properties.

First we discuss changes in the initial mass distribution of the
satellite (making it more or less concentrated). This will modify the
energy distribution of the particles stripped from the satellite, and
hence the rate at which the debris wrap around the orbit.  As
described by \citet{kvj98}, the spread in energy of the debris is
proportional to the ratio of the tidal radius of the satellite to the
radius where particles are stripped. If we assume, as we have done
above, that the satellite disrupts at pericenter, a satellite with
less central concentration (larger tidal radius) will produce debris
with a larger energy spread. This will mean that it will disperse more
rapidly. In reality, a lower concentration satellite (of the same
mass) will disperse at a larger Galactocentric radius, leading to a
smaller energy spread, but an increased spread in angular momentum.
Since the apocenter of the debris is governed by the energy spread and
the pericenter by the angular momentum spread, this will lead to a
larger spread between wraps at apocenter- or pericenter depending on where
the debris is stripped. To first order, these changes will not affect the
detection probabilities.

If the satellite has a larger mass, our assumptions of no self-gravity
or dynamical friction will be less accurate, and it is necessary to
consider the N-body approach. One of the basic differences between
our technique and the full N-body treatment is that our satellites
become unbound immediately, while the N-body satellite have particles
stripped on successive passages. This can be seen in Figure
\ref{jsabound} for orbit 1039.  However, if we consider the case of a
satellite with mass $10^8 M_\odot$ that loses 10\% of its mass on each
pericenter passage, this resembles our simulation with ten $10^7
M_\odot$ satellites becoming totally unbound at pericenter. So, to the
extent that we can ignore self-gravity and dynamical friction, the
detection probability will scale linearly with satellite
mass. Dynamical friction will actually make this approximation better,
as the orbit will be different on each passage where debris are
produced.

\section{Summary}

We have shown that it is possible to efficiently detect the remains of
accreted satellites via their velocity signature. For example, a
{\it single} observation of 100 halo star velocities in a high-latitude
field yields a detection probability of order 1\% for a single 10$^7
M_{\odot} $ satellite against a well-mixed velocity background.
Detections remain possible down to levels where the satellite debris
contributes only a few percent of the stellar density in the field.
Somewhat surprisingly, the detection efficiencies are not strongly
dependent on the age of the total debris.  The competing effects of
the debris spreading over a larger fraction of the sky with time,
and the decreasing impact of its velocity signature on the histogram
due to the decreasing density of debris, approximately cancel. For
orbits with small mean radii, the detections are further compromised
by the occurrence along the line of sight of debris on multiple wraps
of the orbit. This leads to multiple velocity peaks and, in the
limiting case, the distribution of velocities is close to the
underlying well-mixed distribution.

The velocity signatures of the detected satellite debris show a broad
range of properties which bear little relationship to the expectations
of narrow velocity peaks that result from phase space
conservation. Projection effects and the presence of particles from
multiple wraps of the orbit dominate the observed velocity
distributions. Debris from satellites with mean Galactocentric radius
greater than 40  kpc
typically have velocity dispersions of $\simeq 5$ km/s almost
independent of the age of the debris. Satellites with mean
Galactocentric radius more than
25  kpc have widths of tens of km/s at late times.

The detection probabilities derived for a single satellite seen
against a well-mixed halo should generalize to cases where debris from
multiple satellites is present in the field. Because the detection
probabilities are dominated by the small filling factor of most of the
orbits of interest, the probability of detecting a velocity signal in
any field will initially scale nearly linearly with the number of
satellites accreted up to a mass fraction of 10 to 50 percent,
depending on the distribution of orbital radii.
The debris from satellites with different initial conditions, higher
stellar mass, or different density profiles will have similar
detection probabilities (within a factor of 2 for orbits with larger
mean radii), but timescales will be different. 

It is important to invest the extra telescope time to obtain
velocities as part of the survey, otherwise only the most striking
examples (with large spatial overdensities) will be identified. Older
accretion events or those from smaller satellites will never be found.
In contrast to the $\sim$4 times spatial overdensity of the recent
Sloan result \citep{sloanrr}, velocity information allows the
identification of debris with spatial overdensities a factor of 100
less.

\acknowledgements PH wishes to thank Willy Benz for helpful
discussions at the early stages of this work, Chris Mihos and Matthias
Steinmetz for comments
which improved an early draft, and also Rob Kennicutt and
Jim Liebert for their continued support and encouragement.  
PH also thanks the Director and staff of the CWRU astronomy department
for their generous provision of facilities while the paper was written.
This work
was supported by NSF grants AST 96-19490 to HLM, AST 95-28367,
AST96-19632 and AST98-20608 to MM, and AST 96-19524 to EWO.

\pagebreak


\begin{figure}  
\caption{The apocenter, period and pericenter of the 180 satellite
orbits are shown plotted against the mean radius of the orbit. For
quick reference, an orbit's apocenter is approximately 1.4 times the
mean radius and the period in Gyr is approximately 2\% of the mean
radius in kpc. The cutoff in pericenter at the top of the lower panel
is set by the selection criterion that the satellite orbit needs to
penetrate the survey volume, and a circular orbit (whose apocenter is
equal to its pericenter) defines the diagonal cutoff at the left.}
\label{orbit_properties} 
\end{figure}

\begin{figure}  
\caption{The projection onto the meridional plane and the Z-projection
of the satellite debris at 10 Gyr, with and without self-gravity.  In
each of the figures the ordinate and abscissa have the same scale, as
indicated on the ordinate.  Due to the unequal abscissa scales for
different orbits, only their zero points are shown.}
\label{jsax}
\end{figure}

\begin{figure}  
\caption{The $V_z$ and $V_y$ projections of the satellite debris at 10
Gyr, with and without self-gravity. See text for details.}
\label{jsav}
\end{figure}

\begin{figure}
\caption{The bound fraction of the satellite mass as a function of time
for the N-body models.}
\label{jsabound}  
\end{figure}

\begin{figure}  
\caption{Two snapshots at 1 and 10 Gyr, heavy and light points
respectively, of the evolution of three disrupted satellites on orbits
where phase space mixing is relatively slow.  Left two panels show
spatial X-Y and Z-Y projections, right panel shows radial velocity
with respect to the Galactic center (RV$_{gc}$) versus galactocentric
distance.  The three orbits were chosen to have decreasing mean
radius, pericenter and apocenter (and hence increasing spatial mixing)
as we move from top to bottom.  Structure in velocities remains
clearly visible in the right hand panels despite the increase in
spatial mixing seen in the left hand panels.  Only a subsample of 2\%
of the 200,000 particles in the models are plotted for clarity.}
\label{xyzgalclean} 
\end{figure}

\begin{figure}
\caption{ Similar to Figure \ref{xyzgalclean}, but for satellites on
three orbits with much smaller mean radii, where spatial mixing is
more rapid.  In all but the bottom row (orbit 1082) the particles on
different orbital wraps remain clearly detectable in velocities in the
RH panels despite the spatial mixing.}
\label{xyzgalmessy}   
\end{figure}

\begin{figure}  
\caption{The same orbits as seen in Figure \ref{xyzgalclean} at 1 and
10 Gyr (heavy and light points respectively) are plotted, but from a
heliocentric perspective. Only those particles within 30 kpc of the
Sun are shown. The left hand panel shows the distribution of particles
over the sky in $l$ \& $b$.  The right hand panel shows radial
velocity and distance with respect to the Sun (RV$_{\odot}$,R$_{\odot}$).
The middle panel shows $l$ \& $R_{\odot}$, providing a link between the
spatial structure seen in $l$ \& $b$ and the velocity structure in the
right hand panels.  The fraction of particles sampled from the model
has been increased by a factor of 5 from Figure \ref{xyzgalclean}, to
10\%, so that details of the orbits are still visible. Note that the
appearance of the orbits are strongly influenced by their relationship
to the Sun's position.}
\label{lbsunclean}
\end{figure}

\begin{figure}
\caption{ Similar to Figure \ref{lbsunclean} but for the orbits with
small mean radii seen in Figure \ref{xyzgalmessy}.  The increased
density of particles compared with Figure \ref{lbsunclean} is due to
most of the orbits falling within 30 kpc of the Sun.}
\label{lbsunmessy}
\end{figure}

\begin{figure}  
\caption{The spatial and velocity distribution of the particles from
satellite 1082 seen at age 1 Gyr.  The upper panel shows the
appearance of the particles on the sky, and the middle panel is a
longitude velocity plot. At this age the satellite particles have
already wrapped three times round the orbit. However, 95\% of the
particles are concentrated between two consecutive apocenters.  The
velocity histograms in the lower panels show the heliocentric
velocities of particles from fields 2 degrees on a side. The $(l,b)$
coordinates are indicated at the top right of each histogram. The
longitude of the fields in the upper two panels is indicated by the
dotted line.}
\label{rvlb1messy}
\end{figure}

\begin{figure}
\caption{Similar to Figure \ref{rvlb1messy} with the satellite 1082
now seen at an age of 10 Gyr. The broad range of velocities extending
over approximately $\pm $ 200 km/sec is due to particles being
observed over a large fraction of the orbit on many wraps along each
line of sight. }
\label{rvlb2messy}  
\end{figure}

\begin{figure}
\caption{Similar to Figure \ref{rvlb2messy} but for satellite 1197
at age 4 Gyr. }
\label{rvlb1clean}  
\end{figure}

\begin{figure}  
\caption{Similar to Figure \ref{rvlb1clean} but for satellite 1197 at
age 10 Gyr. Note the lack of velocities near 0 km/s and the patchy spatial
distribution --- this is due to the distance limits of $R_{\odot} < 30$
kpc. These extreme velocities make such streams easy to detect,
despite their low spatial density. }
\label{rvlb2clean} 
\end{figure}

\begin{figure}
\caption{The fields used for the observations of the models are shown
on an equal area polar projection. The point size approximately
matches the field size of 1 degree.}
\label{fieldlb} 
\end{figure}
\clearpage

\begin{figure}  
\caption{A sample of the velocity and distance distributions of
particles that lead to detections based on the observed velocities.
The unshaded histogram gives the combined distribution of satellite
plus smooth halo particles observed in the field.  The shaded
histogram shows the distribution of the satellite particles only.  The
upper and lower halves of the figure show the magnitude and velocity
distributions of the satellite particles in fields with detections.
Four examples are shown for each satellite; the $l,b$ coordinates of
the field are in the upper left of all histograms.  The $P$-value, the
probability of the detection occurring at random, is shown in the upper
right of the velocity histograms.  The age of the tidal debris in Gyr
is shown in the upper right of the magnitude histograms.  The
velocities include observational one sigma errors of 20 km/s and the
magnitudes include the 0.35 sigma distribution in the absolute
magnitude of halo turnoff stars.}
\label{hist32} 
\end{figure}

\begin{figure}  
\caption{The lower panel shows the relationship between energy and
velocity (RV$_{gc}$) of particles from two of the detections of 
Figure \ref{hist32}. The middle panel shows the observational quantities
velocity with respect to the Sun (RV$_{Sun}$) and distance (R$_{\odot}$)
while the upper panel shows the effect of observational errors by
plotting apparent magnitude ($V_{mag}$) with the 0.35 mag spread in
luminosity at the turnoff added, versus velocity degraded by the 20
km/s observational error.  The satellite orbit and $l,b$ coordinates
of the field are shown at the top of the two columns.  The small
points in the lower panels show the clear sorting of particles in
energy due to phase space conservation.  The crosses identify the
satellite particles within the field. Gaps between crosses are
primarily due to the width of each wrap which is smaller than the
field size.  The energy spread of the particles from both orbits is
only a few km/s$^{2}$ (or 0.2 percent). Despite this small range of
energy, the particle velocities have a range of more than 30 km/s in
each case.}
\label{erm} 
\end{figure}

\begin{figure}  
\caption{Components of the probability of detecting the debris from a
single satellite on a given initial orbit seen against a dynamically
well mixed halo are plotted against the mean radius of the initial
orbit of each satellite. The upper panel shows the accessible
percentage of the orbital period, set by the survey radius of 20 kpc.
The probability of 5 or more satellite particles being present in a
field, the minimum required for a detection is shown in the second
panel from the top. The next panel shows the conditional probability
of detecting a satellite debris in a field when five particles are
present.  The lower panel shows the final detection probability and is
the product of the two components above.  The detection probabilities
are the averages over the five 0.5 Gyr time steps from 5.5 to 7.5 Gyr
and averaged over the 61 fields for which the models were observed.}
\label{pcondc} 
\end{figure}

\begin{figure}  
\caption{The variation with time of the probability of detecting the
debris from a single satellite seen against a dynamically well mixed
halo in a single field. From top to bottom the panels shows the
detection probability at 1.5, 4, 6.5 \& 9 Gyr after the first
perigalactic passage of the satellite.  The probabilities are the
average over the 5 surrounding time steps of 0.5 Gyr each.}
\label{pagead} 
\end{figure}

\begin{figure}  
\caption{The detection probability per field as a function of time
since satellite destruction is shown by the solid line for the six
satellites in figures \ref{xyzgalclean} \& \ref{xyzgalmessy}.  The
satellites shown, 1183, 1103, 1197, 1289, 6866, \& 1082 from top to
bottom are in decreasing order of mean radii. The dotted line is the
probability per field of five or more stars from the satellites are
present in the field and is scaled by the right hand axis.}
\label{pfd_cy20} 
\end{figure}

\begin{figure}  
\caption{ Comparison of average detection probabilities when all, 100
or 50 velocities are ``observed'' in each field as a function of mean
radius of the satellite orbit}
\label{prob3_cy20} 
\end{figure}

\begin{figure}  
\caption{ Similar to figure \ref{prob3_cy20} but with the survey
radius reduced to 12kpc (from the 20 kpc used previously).}
\label{prob3_cy12} 
\end{figure}


\begin{thebibliography}{DUM}
  
\bibitem[Bahcall and Casertano(1986)]{bc86} Bahcall, J.N. \&
  Casertano, S. 1986, \apj, 308, 347
  
\bibitem[Beers et al.(2000)]{beers2000} Beers, T. C., Chiba, M.,
  Yoshii, Y., Platais, I., Hanson, R. B., Fuchs, B. and Rossi, S.
  2000, \aj, in press (June 2000)

\bibitem[Binney and Tremaine(1987)]{bat87} Binney, J. and Tremaine,
  S. 1987, Galactic Dynamics (Princeton: Princeton University Press)

\bibitem[van den Bosch, Lewis, Lake and Stadel (1999)]{bosch99} 
van den Bosch, F. C., Lewis, G. F., Lake, G. and Stadel, J. 1999, \apj, 
515, 50

\bibitem[Colpi et al.(1999)]{dynf99} Colpi, M., Mayer, L. and
  Governato, F. 1999, \apj, 525, 720
  
\bibitem[Carney and Latham(1986)]{cl86} Carney, B. W. and Latham, D.
  W. 1986, \aj, 92, 60
  
\bibitem[Chiba \& Yoshii(1998)]{cy98} Chiba, M.  \& Yoshii, Y.  1998,
  \aj, 115, 168
  
\bibitem[Cora, Muzzio and Vergne(1997)]{dyn97} Cora, S. A., Muzzio, J.
  C. and Vergne, M. M. 1997, \mnras, 289, 253
  
\bibitem[D'Agostino and Stephens(1986)]{the book} D'Agostino, R.B. \&
  Stephens, M.A. 1986, {\it Goodness of Fit Techniques} (Marcel
  Dekker, New York).  

\bibitem[Dohm-Palmer et al.(2000)]{spag2} Dohm-Palmer, R.C., Mateo,
 H.L.,Olszewski, M., Morrison, E.W., Harding, P., Freeman, K.C. and
Norris, J.E. 2000, \aj, in press.

\bibitem[Edvardssen et al.(1993)]{edv93}
  Edvardsson, B., Andersen, J., Gustafsson, B., Lambert, D. L.,
  Nissen, P.E., \& Tompkin, J. 1993, \aap, 275, 101
  
\bibitem[Eggen et al.(1962)]{els} Eggen, O. J., Lynden-Bell, D. and
  Sandage, A. R. 1962, \apj, 136, 748 
  
\bibitem[Ghigna et al.(1998)]{ghigna98} Ghigna, S., Moore, B., 
Governato, F., Lake, G., Quinn, T. and Stadel, J.  1998, \mnras, 300, 146 

\bibitem[Gilmore, Wyse and Jones (1995)]{gw95} 
Gilmore, G., Wyse, R. F. G. and Jones, B. J. 1995, \aj, 109, 1095 

\bibitem[Harding et al.(2000)]{spagn} Harding, P., Morrison, H.L.,
  Mateo, M., Olszewski, E.W., Dohm-Palmer, R.C., Freeman, K.C. and
  Norris, J.E. 2000, in preparation.
  
\bibitem[Helmi and White(2000)]{helmi2000} Helmi, A. and White, S. D.
  M. 2000, \mnras, submitted (astro-ph/0002482)
  
\bibitem[Helmi and White(1999)]{helmi99} Helmi, A. and White, S. D. M.
  1999, \mnras, 307, 495
  
\bibitem[Helmi et al.(1999)]{helmihip} Helmi, A., White, S. D. M., de
  Zeeuw, P. T. and Zhao, H. 1999, \nat, 402, 53
  
\bibitem[Ibata et al.(1994)]{igi94} Ibata, R. A., Gilmore, G. and
  Irwin, M. J. 1994, \nat, 370, 194
  
\bibitem[Ivezic et al.(2000)]{sloanrr} Ivezic, Z., Goldston, J.,
  Finlator, K. et al. 2000, \aj, in press
  
\bibitem[Hernquist(1990)]{lh90} Hernquist, L. 1990, \apj, 356, 359
  
\bibitem[Jiang and Binney(2000)]{jib2000} Jiang, I. and Binney, J.
  2000, \mnras, 314, 468
  
\bibitem[Johnston et al.(1996)]{jhb96} Johnston, K. V., Hernquist, L.
  and Bolte, M. 1996, \apj, 465, 278
  
\bibitem[Johnston et al.(1995)]{jsh95} Johnston, K. V., Spergel, D.
  N. and Hernquist, L. 1995, \apj, 451, 598
  
\bibitem[Johnston et al.(1998)]{kvj98} Johnston, K. V. 1998, \apj,
  495, 308
  
\bibitem[Kinman, Wirtanen and Janes(1965)]{tdk65} Kinman, 
T.\ D., Wirtanen, C.\ A.\ and Janes, K.\ A.\ 1965, \apjs, 11, 223 
  
\bibitem[Klypin et al.(1999)]{klypin99a} Klypin, A., Kravtsov, A. V.,
  Valenzuela, O. and Prada, F. 1999, \apj, 522, 82
  
\bibitem[Majewski(1992)]{srm92} Majewski, S.R. 1992, \apjs, 78, 87
  
\bibitem[Majewski et al.(1994)]{srm94} Majewski, S.R., Munn, J.A., \&
  Hawley, S.L. 1994, \apj, 427, L37

\bibitem[Majewski et al.(2000)]{srm00} Majewski, S.\ R., Ostheimer, J.\ C., Kunkel, W. E. and Patterson, R. J. 2000, to appear in 
October 2000 issue of The Astronomical Journal.

\bibitem[Mateo(1998)]{mateo_araa} Mateo, M. L. 1998, \araa, 36, 435
 
\bibitem[Miyamoto \& Nagai(1975)]{mn75} Miyamoto, M.  \& Nagai, R.
  1975, \pasj, 27, 533
  
\bibitem[Moore et al.(1999)]{moore99a} Moore, B., Ghigna, S.,
  Governato, F., Lake, G., Quinn, T., Stadel, J. and Tozzi, P. 1999,
  \apjl, 524, L19
  
\bibitem[Morrison(1996)]{hlm96} Morrison, H.L. 1996, in ``Formation of
the Galactic Halo ... Inside and Out'', ASP Conference
  Series, Vol. 92, 1996, Eds H.L. Morrison and A. Sarajedini

\bibitem[Morrison et al.(2000a)]{spag1} Morrison, H.L., Mateo, M.,
Olszewski, E. W., Harding, P., Dohm-Palmer, R. C., Freeman, K.  C.,
Norris, J. E. and Morita, M. 2000, \aj, 119, 2254
  
\bibitem[Morrison et al.(2000b)]{spag4} Morrison, H.L., Olszewski,
E. W., Mateo, M., Norris, J. E., Harding, P., Dohm-Palmer, R. C. and 
Freeman, K.  C., \aj, submitted 
  
\bibitem[Navarro et al(1995)]{nfw95} Navarro J.F., Frenk. C.S., \&
  White 1995, S.D.M 1995
  
\bibitem[Norris(1986)]{jen86} Norris, J. 1986, \apjs, 61, 667

\bibitem[Preston et al.(1991)]{psb91} Preston, G.W., Shectman, S.A. and
Beers, T.C. 1991, \apj, 375, 121
  
\bibitem[Preston et al.(1994)]{gwp94} Preston, G. W., Beers, T. C., \&
  Shectman, S. A. 1994, \aj, 108, 538 (PBS)
  
\bibitem[Royston(1995)]{roy95} Royston, J.P. 1995, Appl. Statist., 44,
  565
  
\bibitem[Saha(1985)]{abi85} Saha, A. 1985, \apj, 289, 310
  
\bibitem[Searle and Zinn(1978)]{sz78} Searle, L. and Zinn, R, 1978,
  \apj, 225, 357
  
\bibitem[Shapiro and Wilk(1965) ]{sw65} Shapiro, S. S. and Wilk, M. B.
  Biometrika, 52, 591
  
\bibitem[Sommer-Larsen and Zhen(1990)]{sl} 
Sommer-Larsen, J.\ and Zhen, C.\ 1990, \mnras, 242, 10 

\bibitem[Steinmetz and Mueller(1994)]{ms94} 
Steinmetz, M.\ and Mueller, E.\ 1994, \aap, 281, L97 

\bibitem[Sun et al.(2001)]{smhw}
Sun, Jiayang, Morrison, H.L., Harding Paul and Woodroofe, Michael
2001, submitted to the Journal of Journal of the American Statistical
Association.

\bibitem[Sun and Woodroofe(1996)]{jiayang}
Sun, Jiayang and Woodroofe, Michael 1996, Journal of Statistical
Planning and Inference, 52, 143

\bibitem[Tremaine(1993)]{tre93} Tremaine, S. 1993, in Back to the
  Galaxy eds. S. S. Holt \& F. Verter, (AIP Conf. Proc. : New York),
  p.599
  
\bibitem[Tremaine(1999)]{tre_geom} Tremaine, S. 1999, \mnras, 307,
  877

\bibitem[Walker et al.(1996)]{walker96} Walker, I. R., Mihos, J. C. and
  Hernquist, L. 1996, \apj, 460, 121

\bibitem[White and Sprigel(2000)]{sdm2000}
White, S. D. M. and Sprigel V. 2000 astro-ph/9911378

\bibitem[Woolley(1978)]{woolley} Woolley, R. 1978, \mnras, 184, 311
  
\bibitem[Yanny et al.(2000)]{yanny2000} Yanny, B., et al.
  2000, \apj, in press
  
\bibitem[Zaritsky et al.(1989)]{dz89} Zaritsky, D., Olszewski, E. W.,
  Schommer, R. A., Peterson, R. C. and Aaronson, M. 1989, \apj, 345,
  759
  
\bibitem[Zhao et al.(1999)]{hz99} Zhao, H., Johnston, K. V.,
  Hernquist, L. and Spergel, D. N. 1999, \aap, 348, L49

\bibitem[Zinn(1993)]{zin93} Zinn, R., 1993, in {The Globular Cluster-
    Galaxy Connection,} ASP Conf Series 48, edited by G. H.~Smith \&
  J. P. Brodie, (ASP, San Francisco), p. 38

\end{thebibliography}
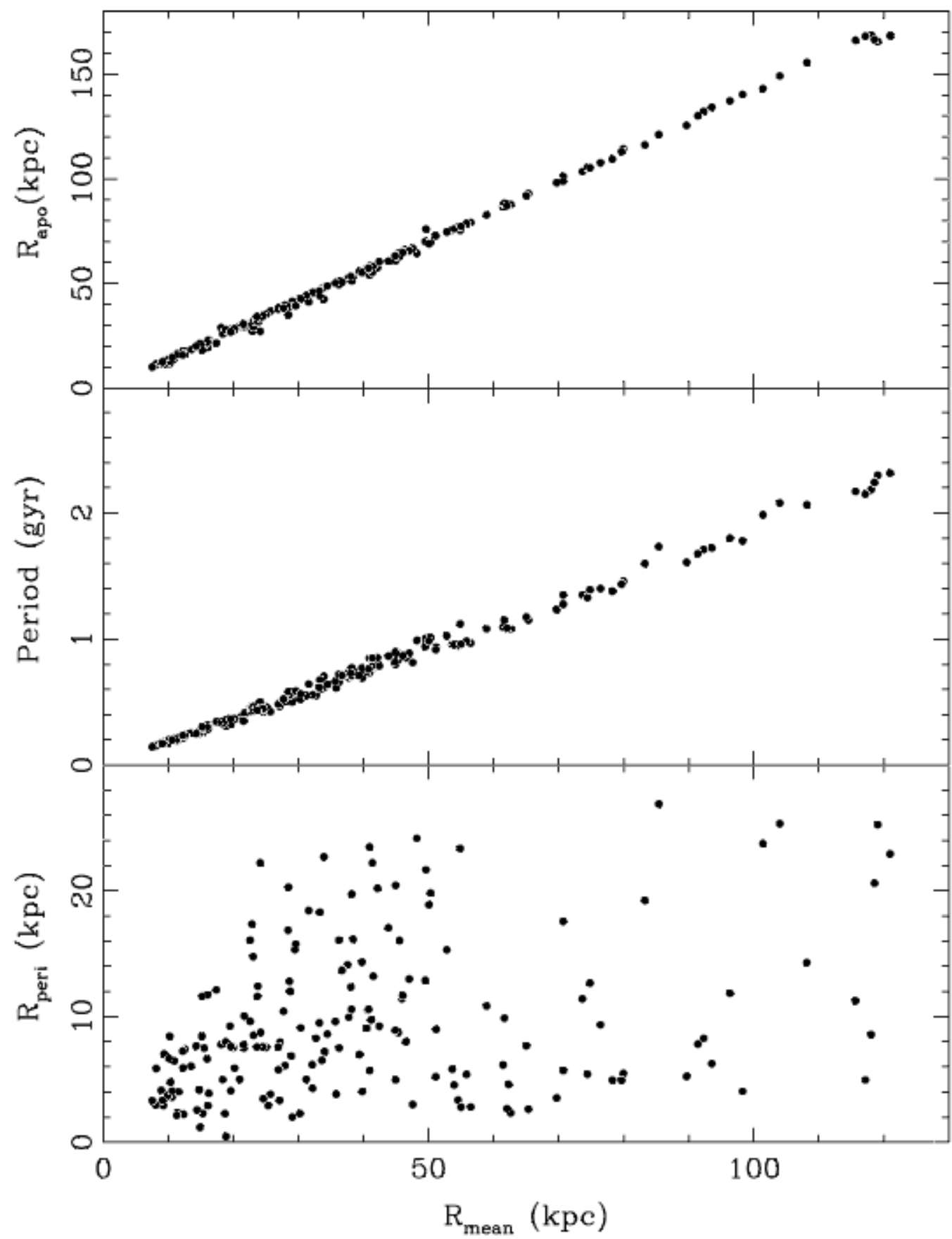
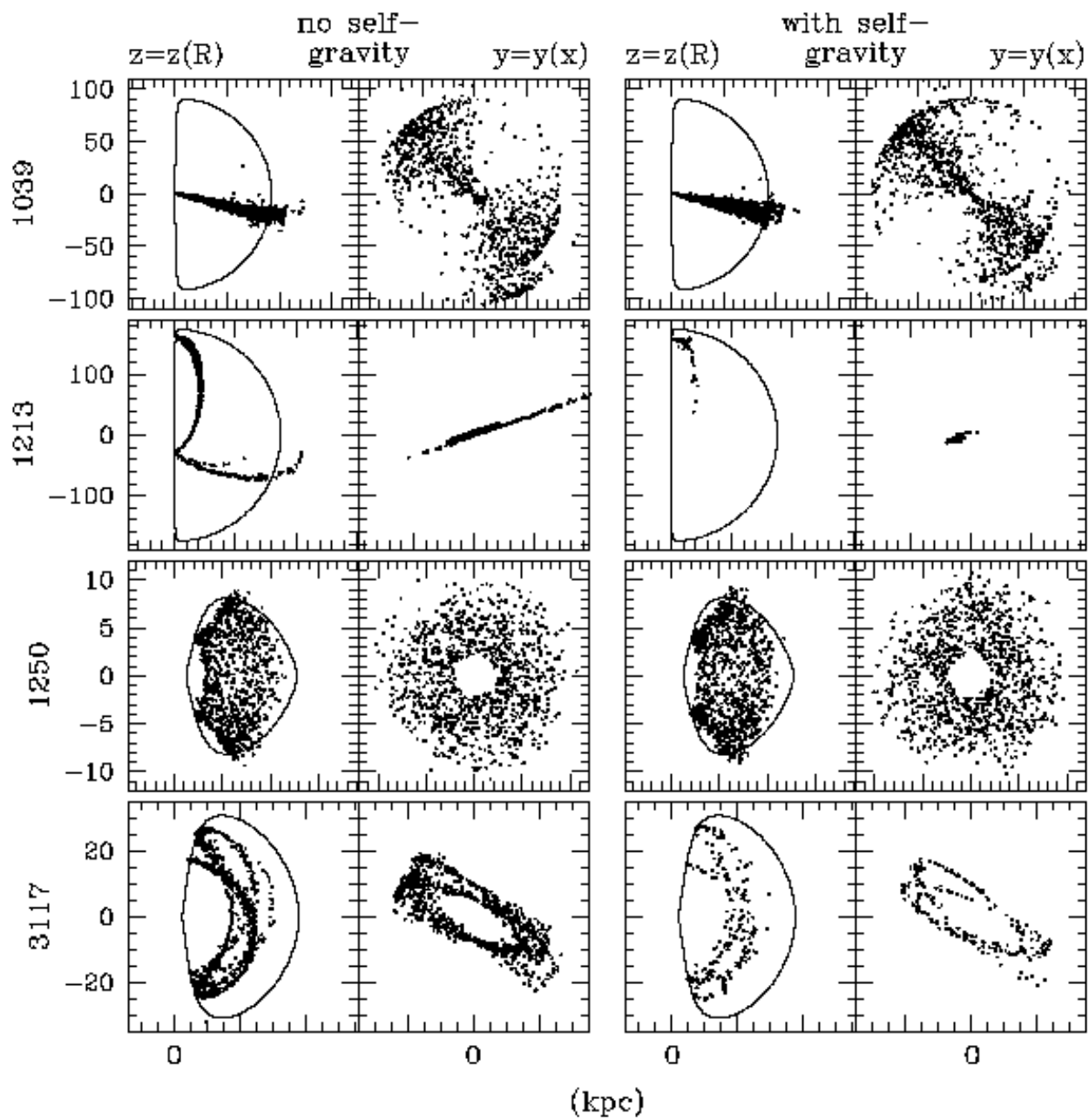
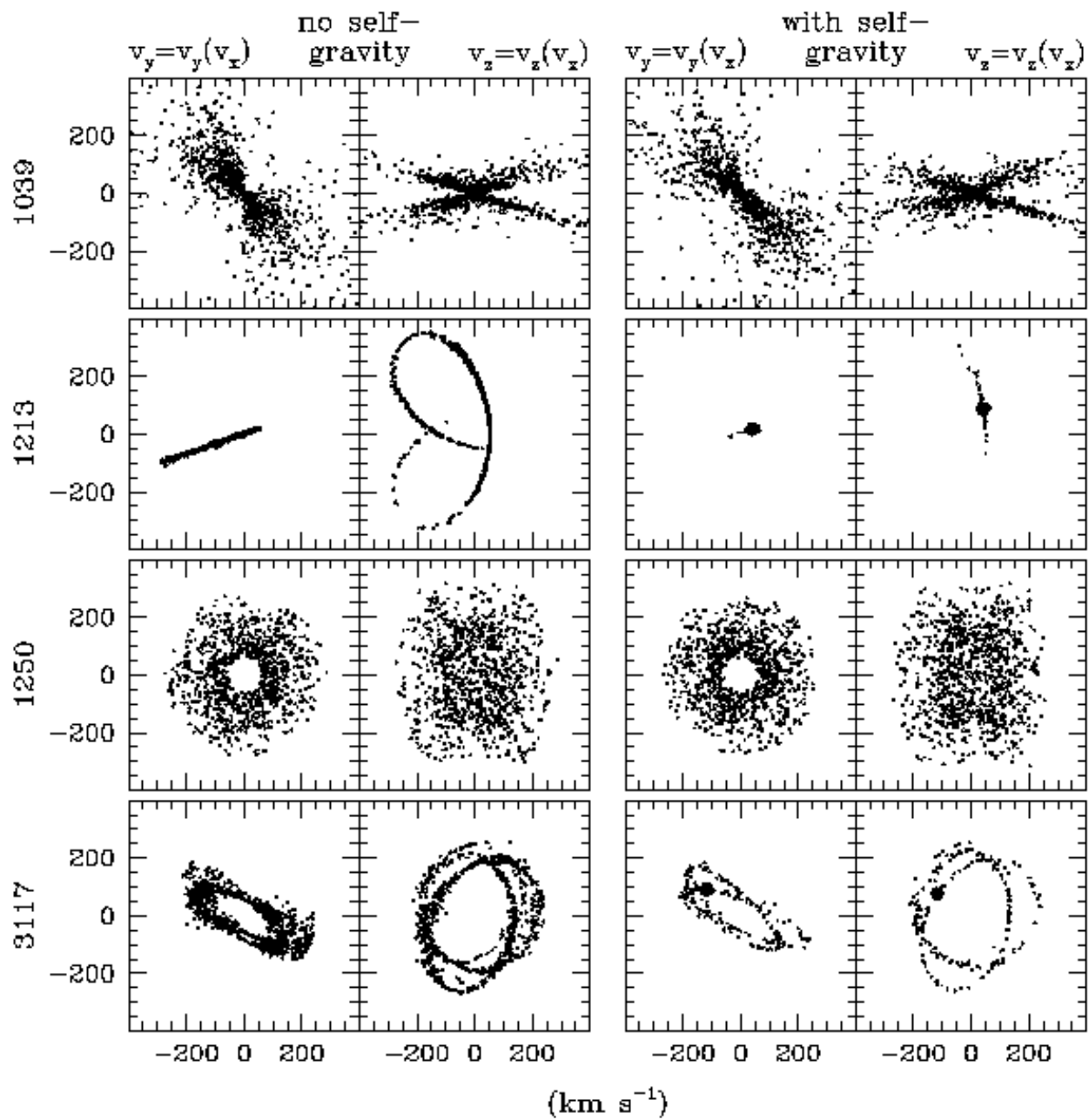
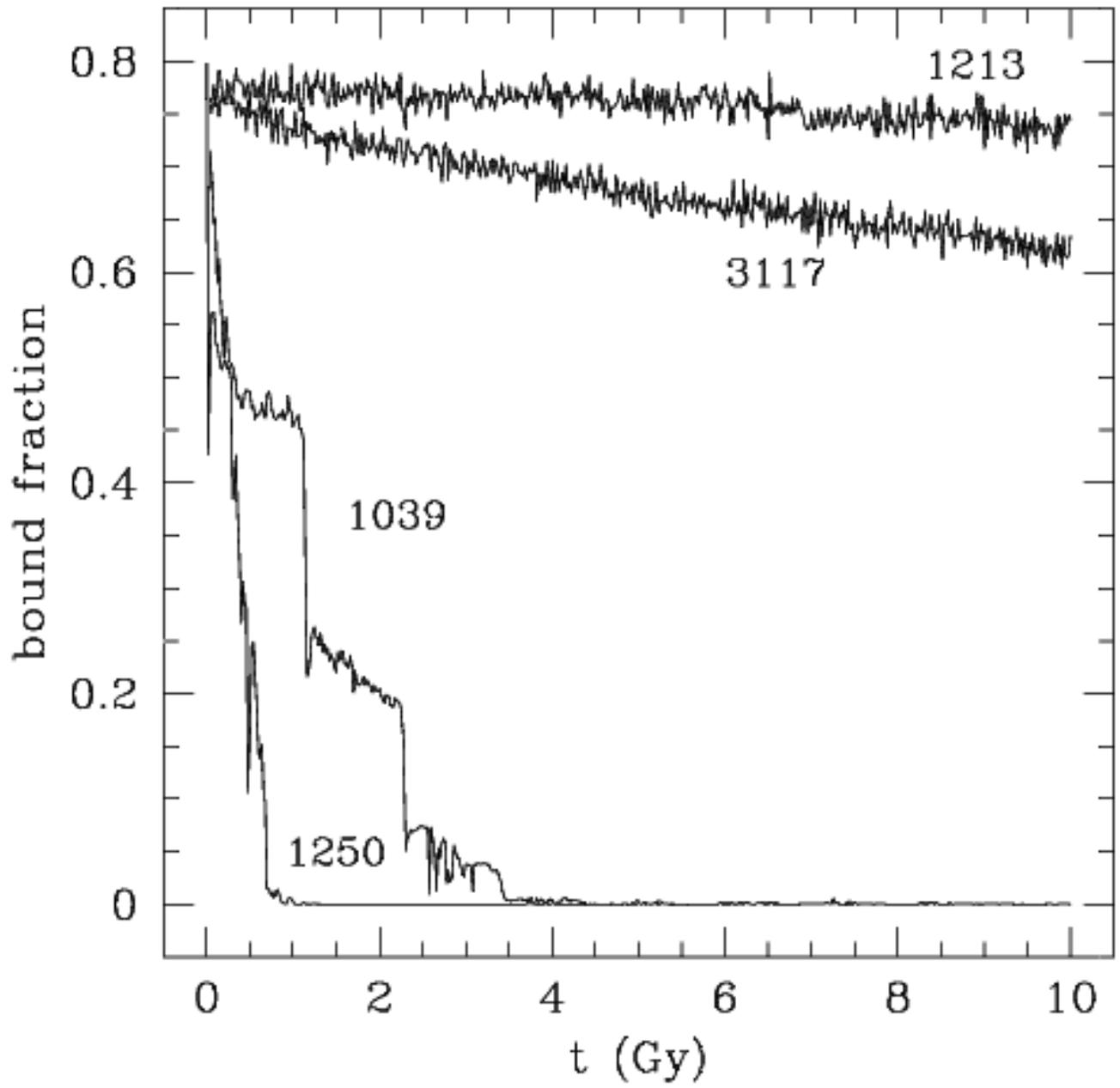
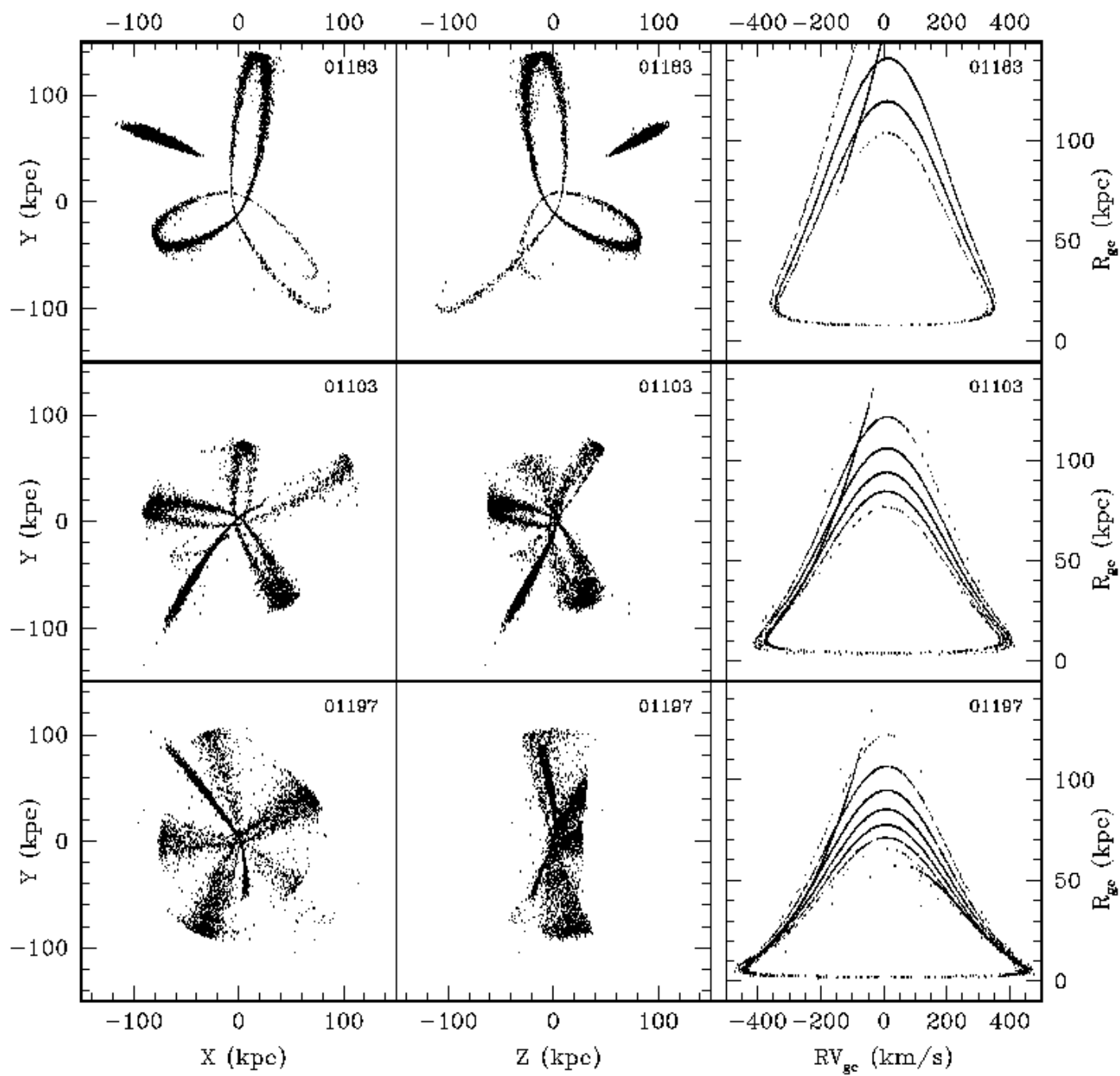
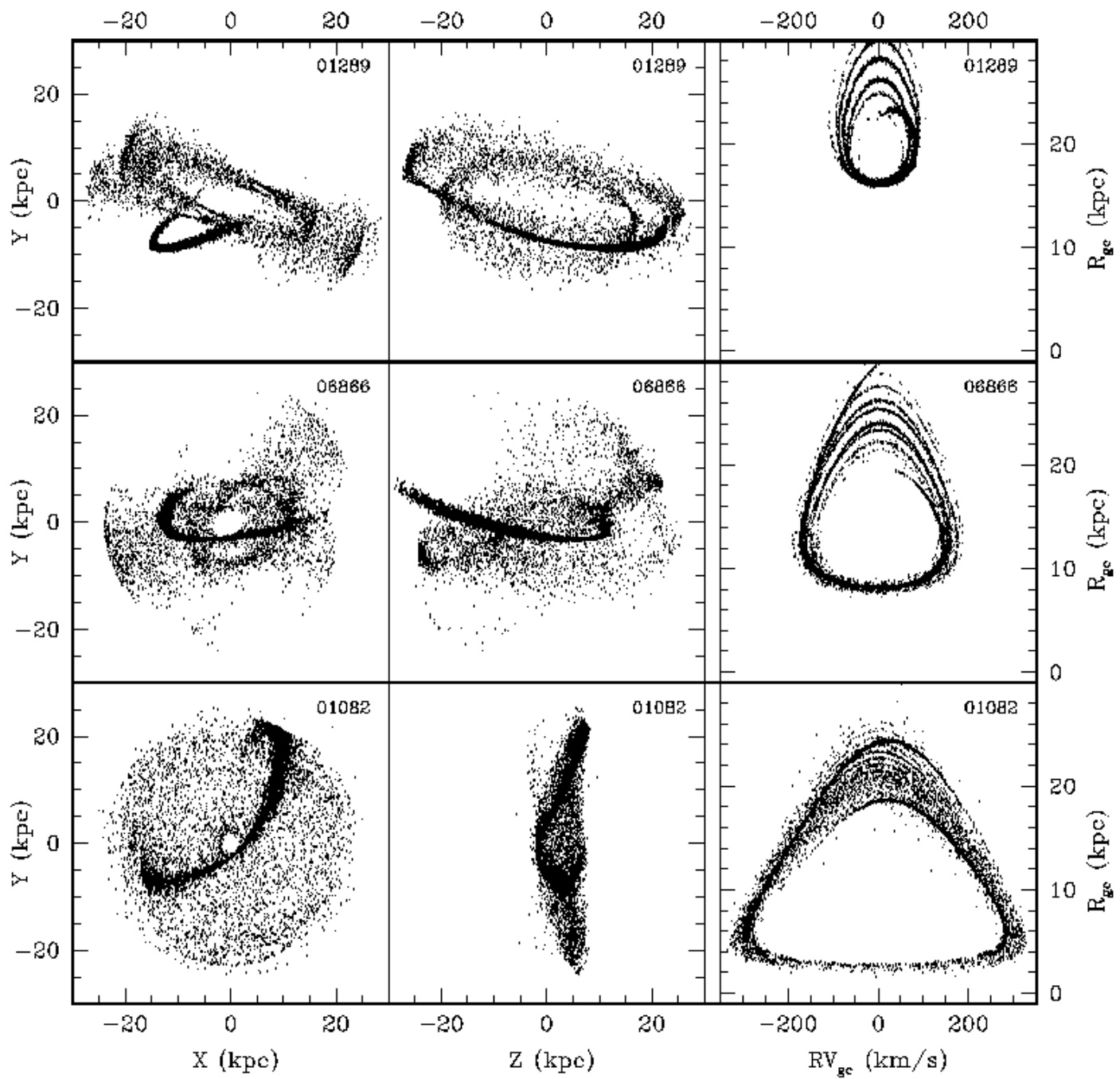
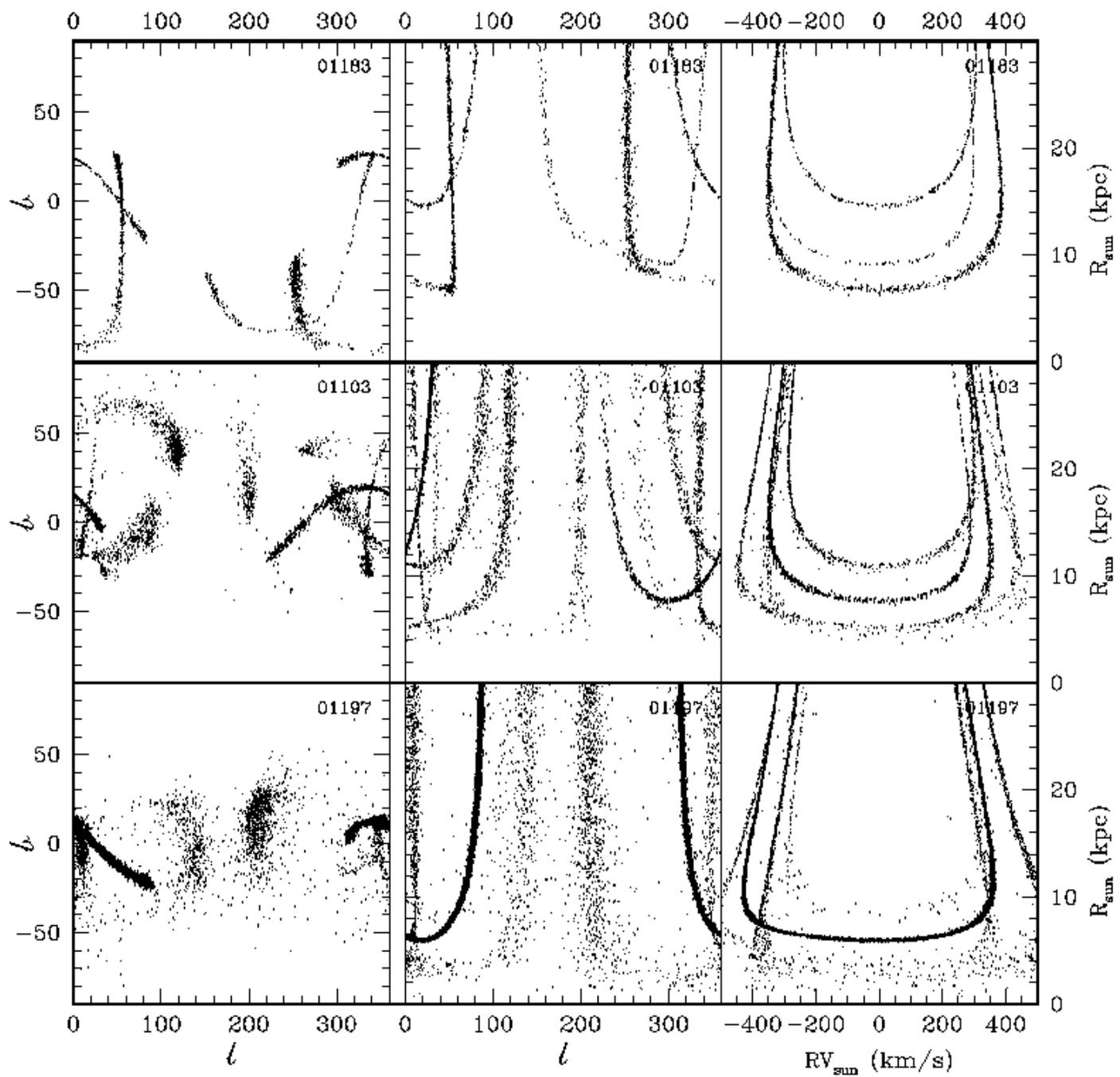
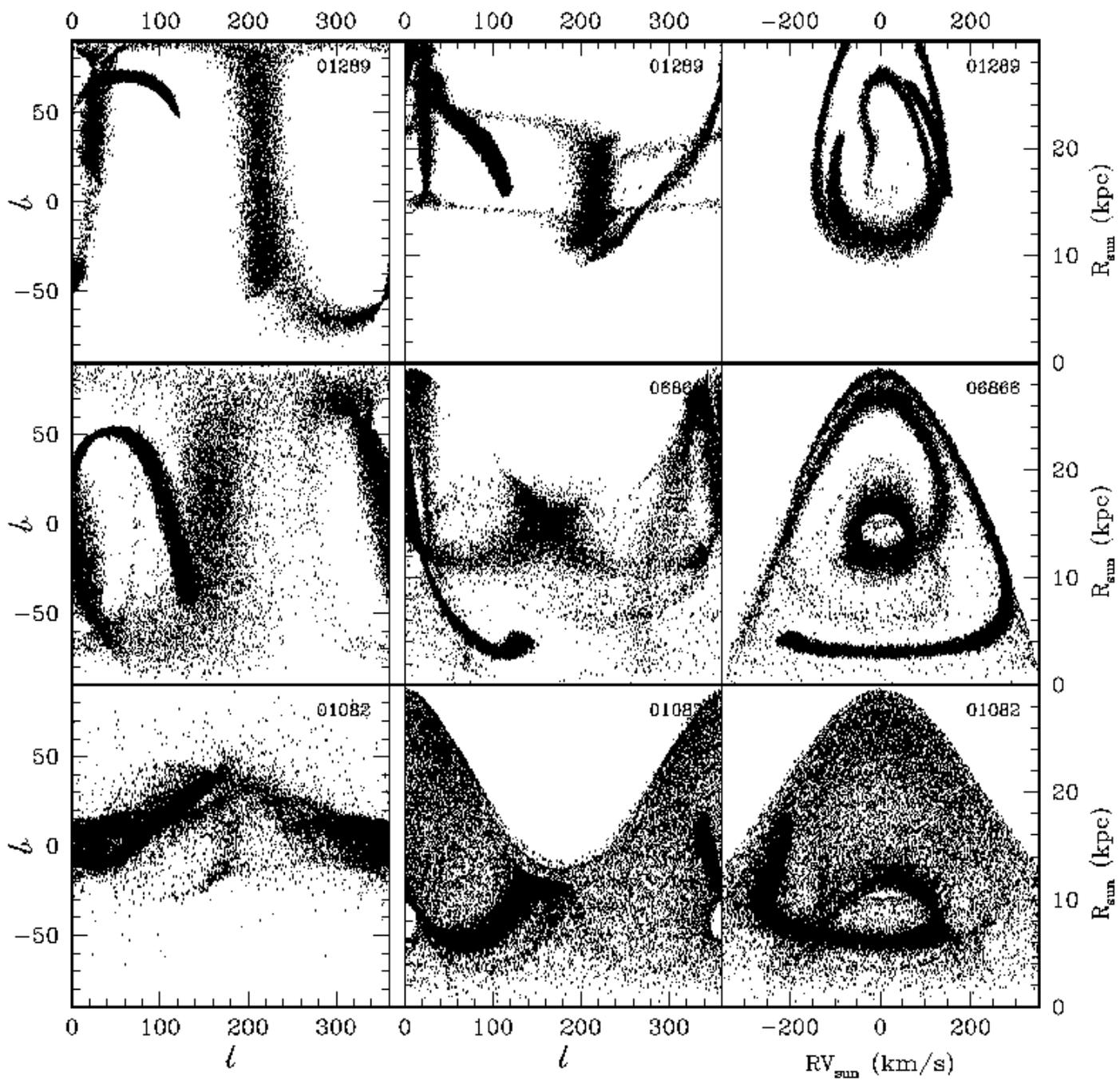
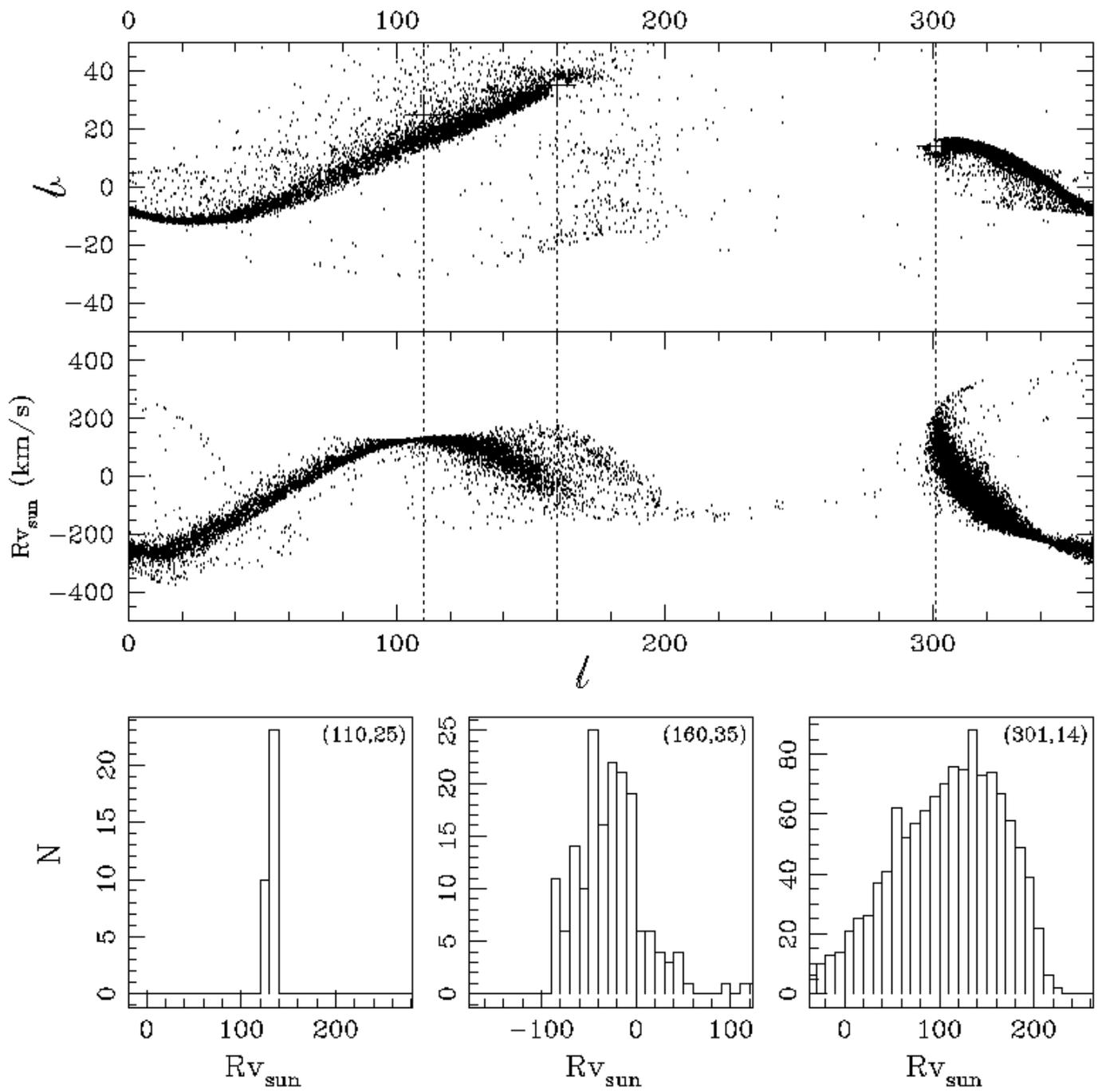
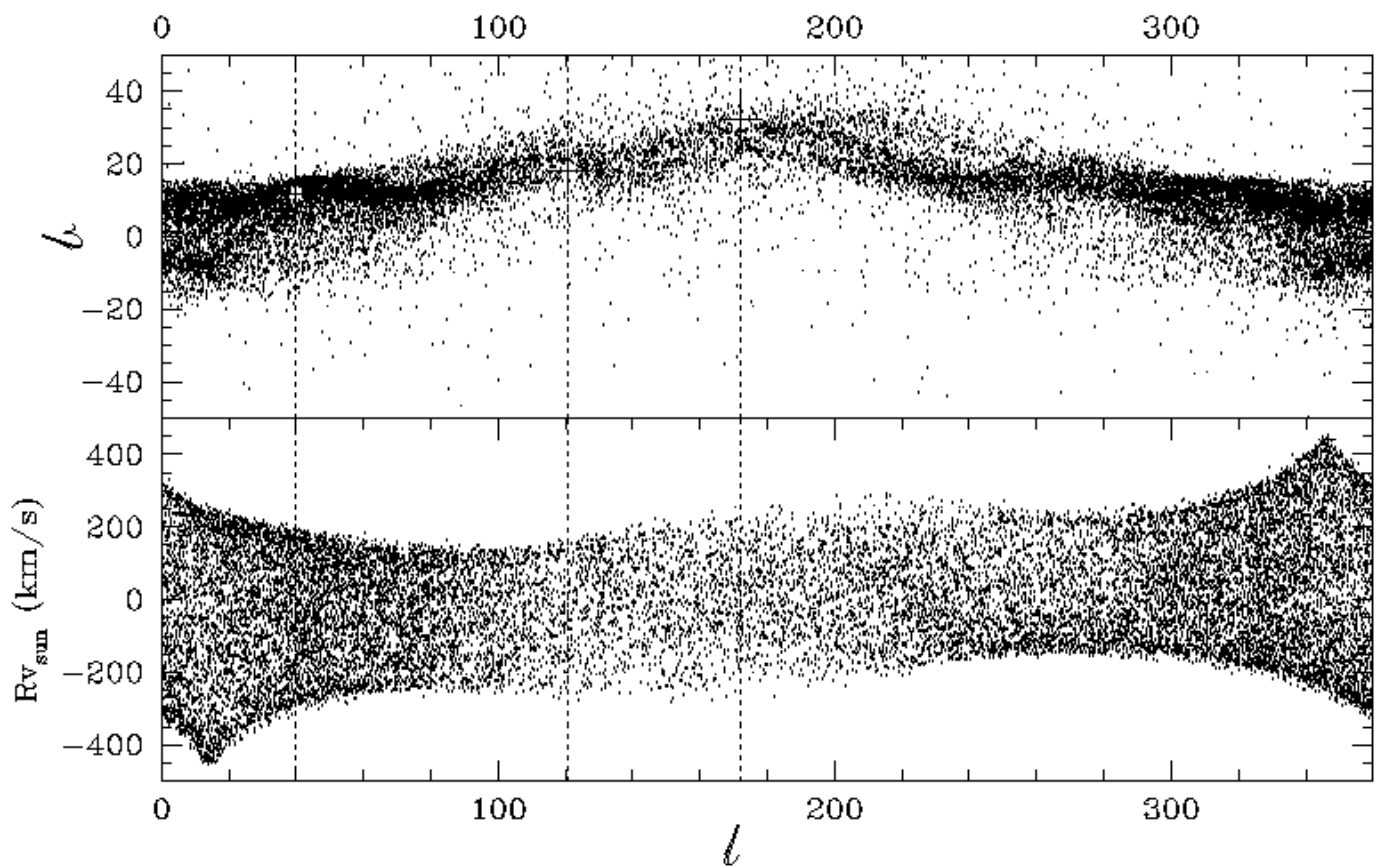
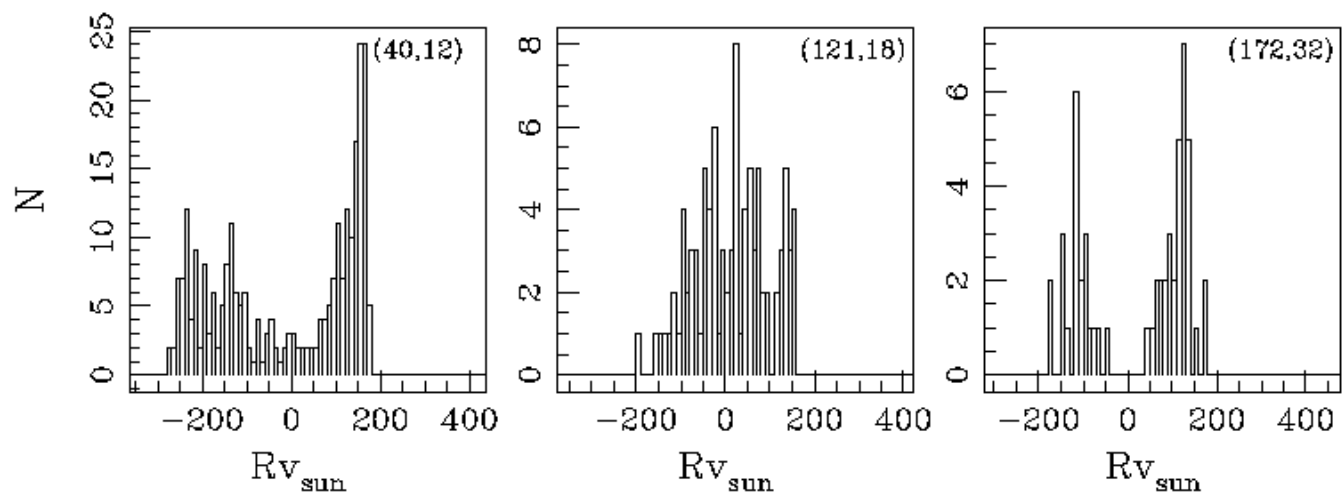
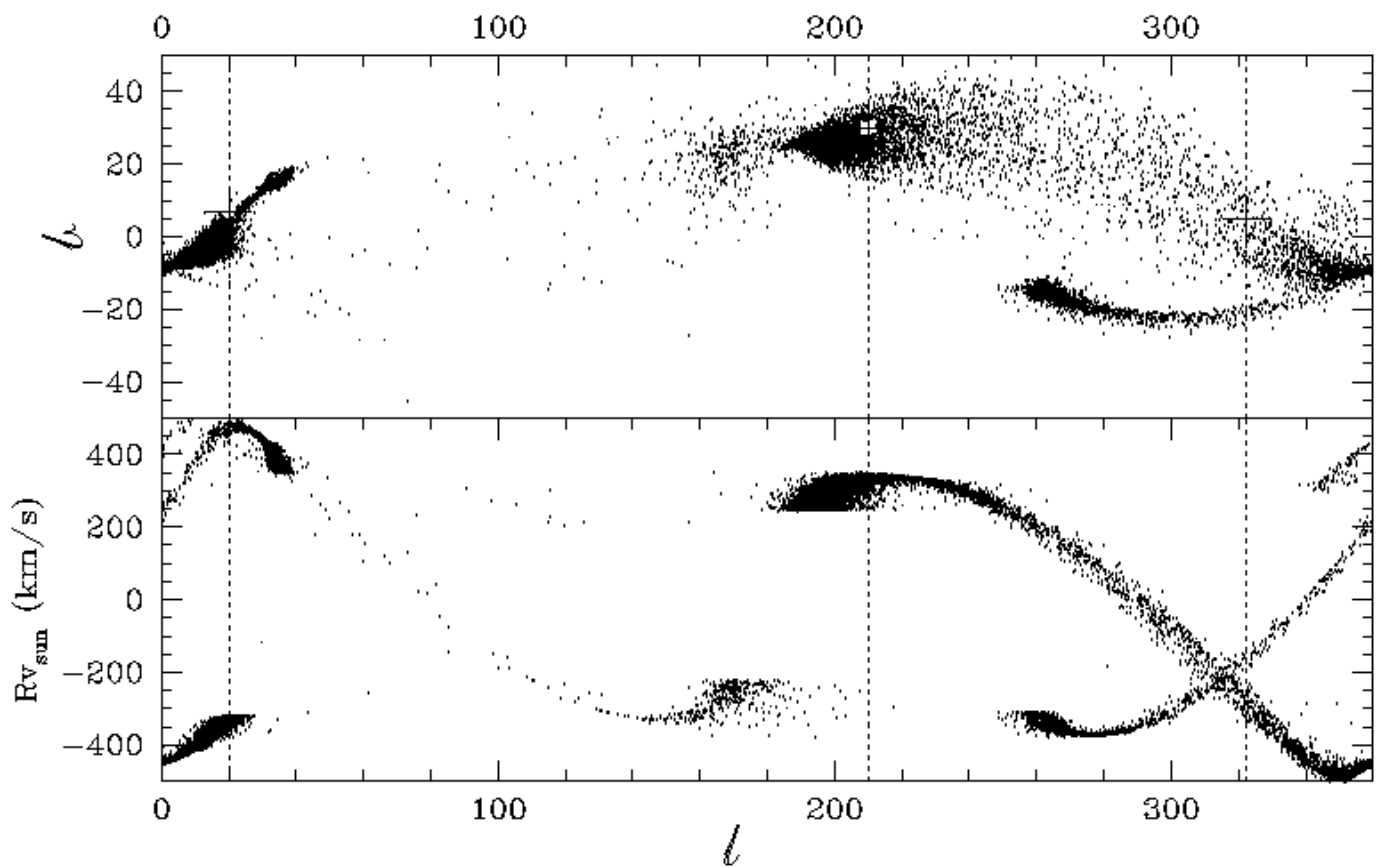
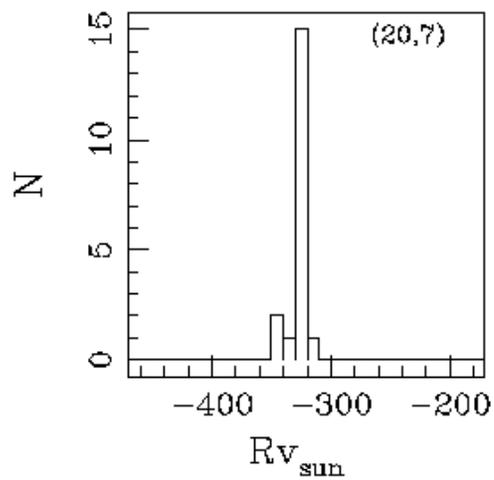
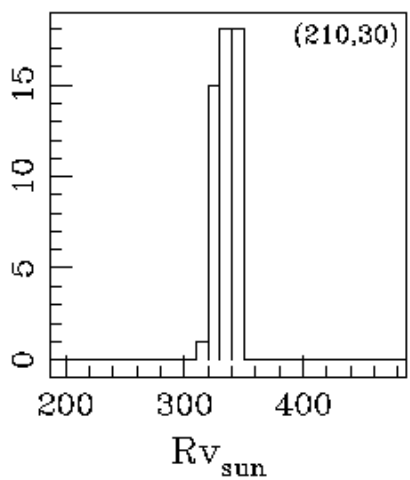
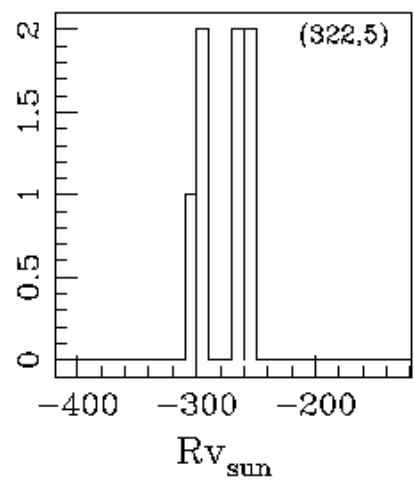
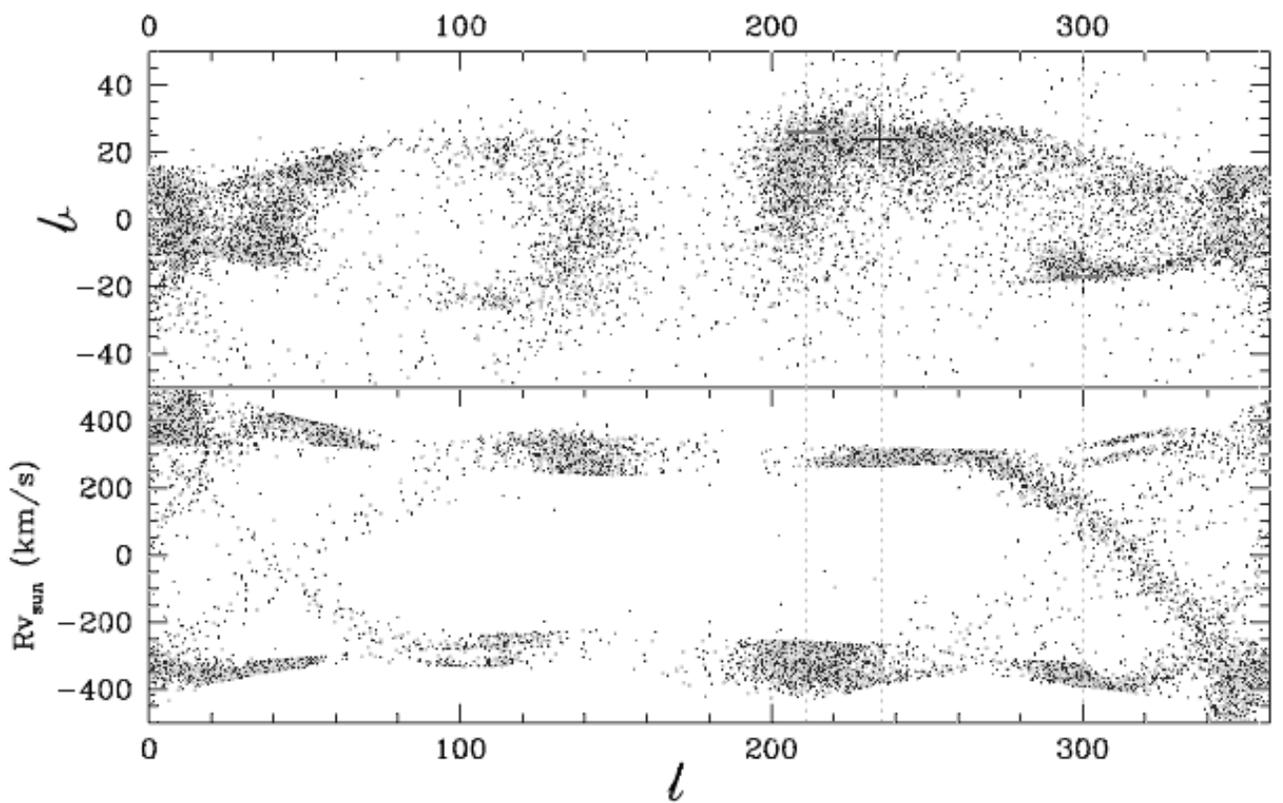
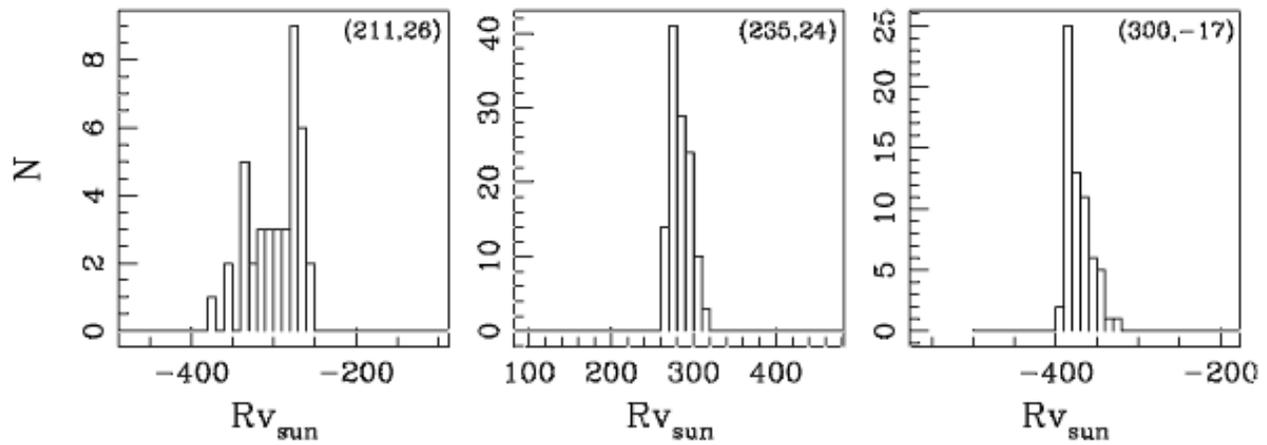
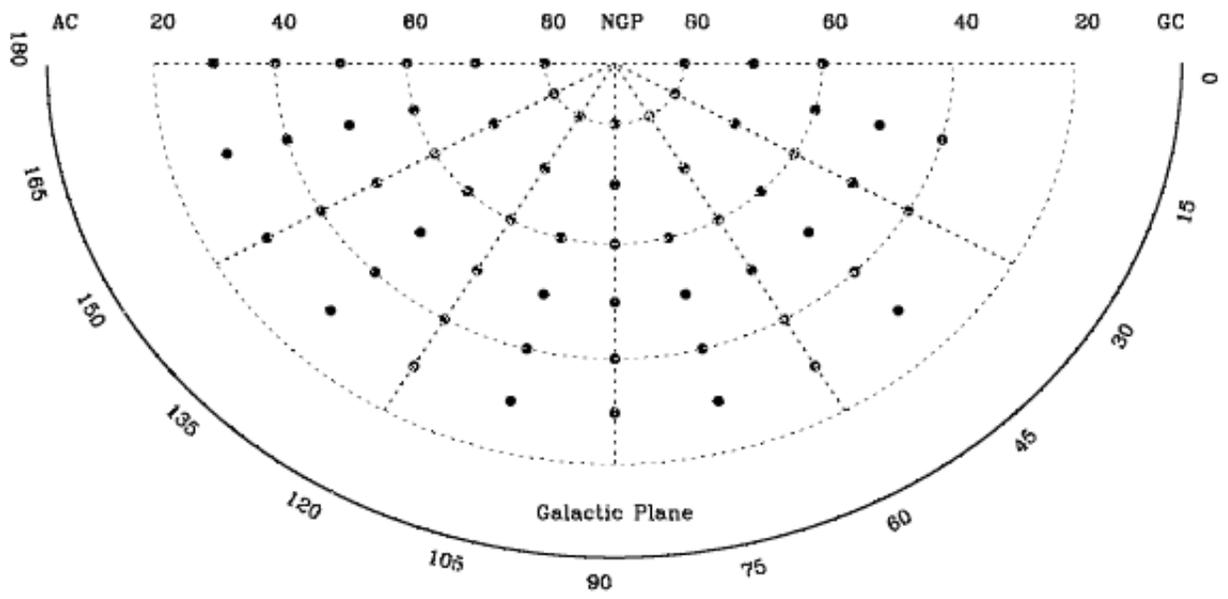
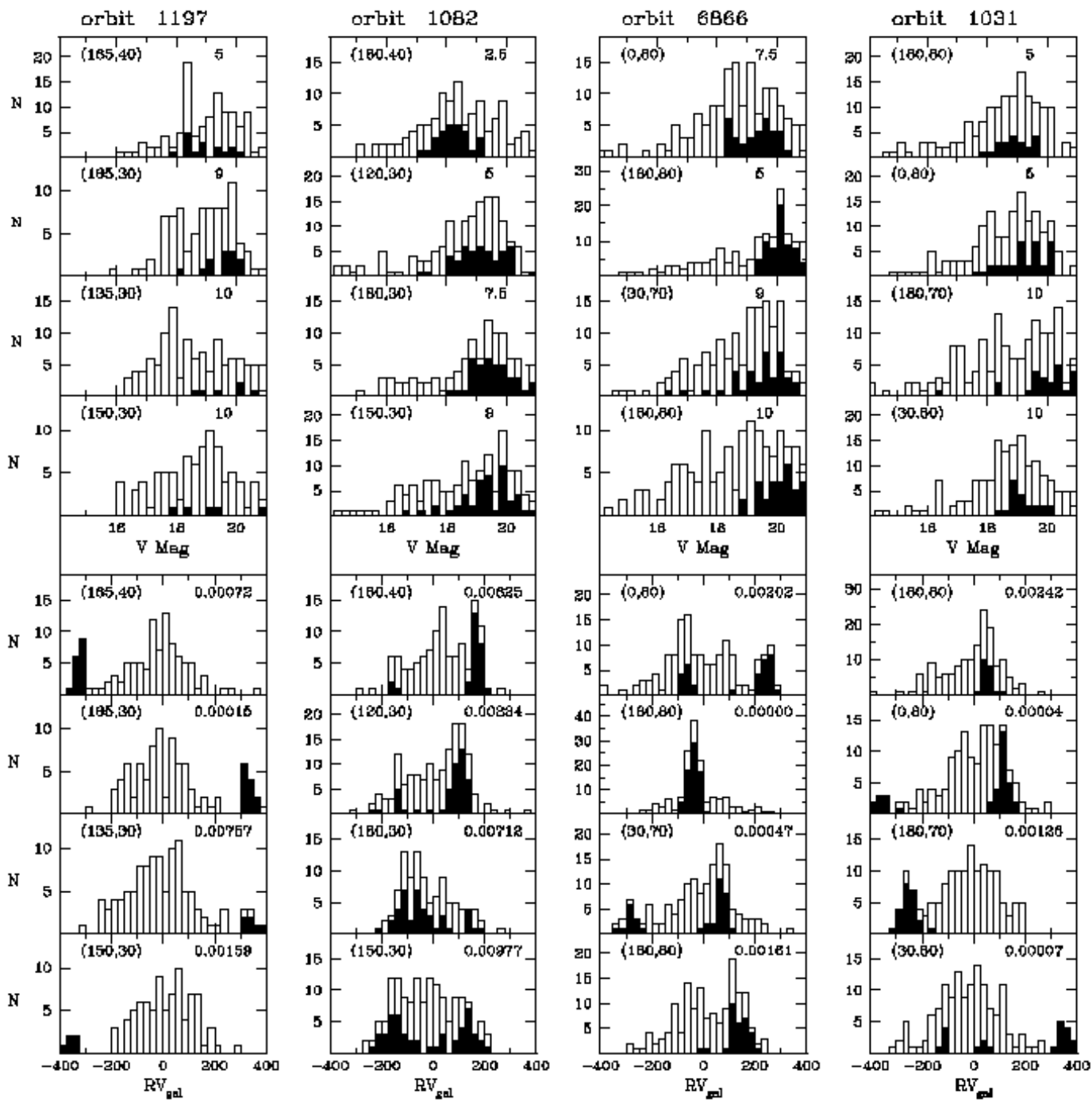
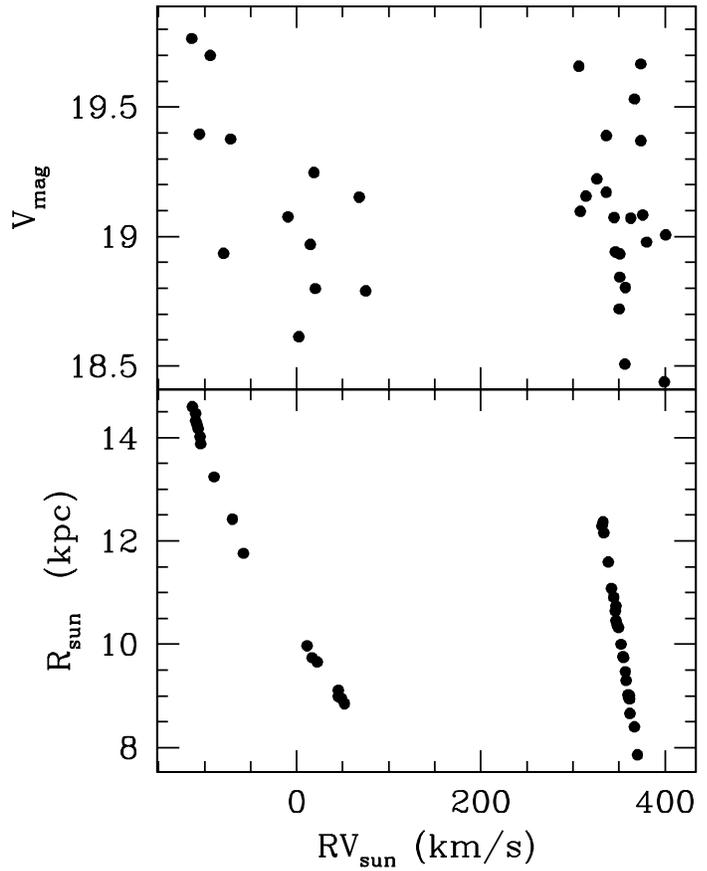
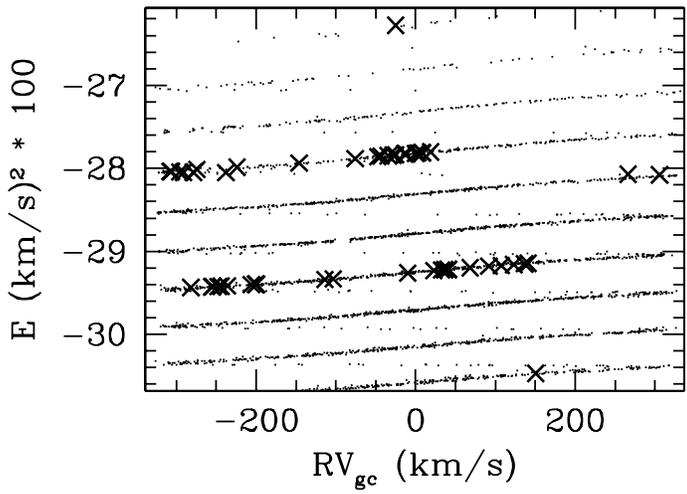
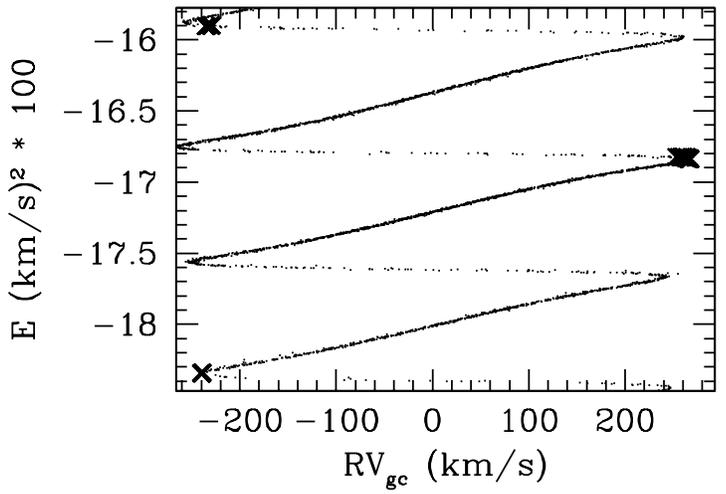
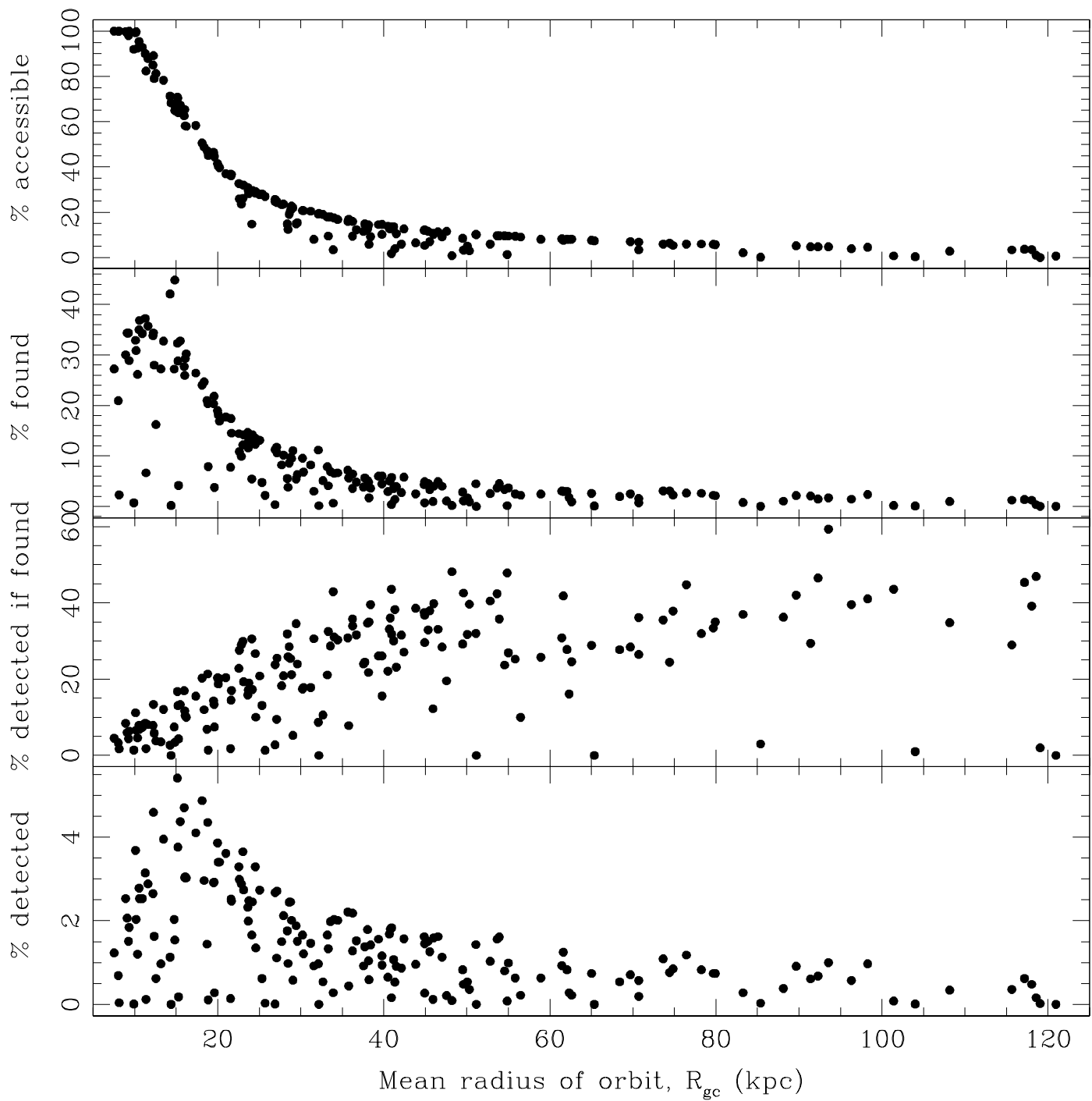
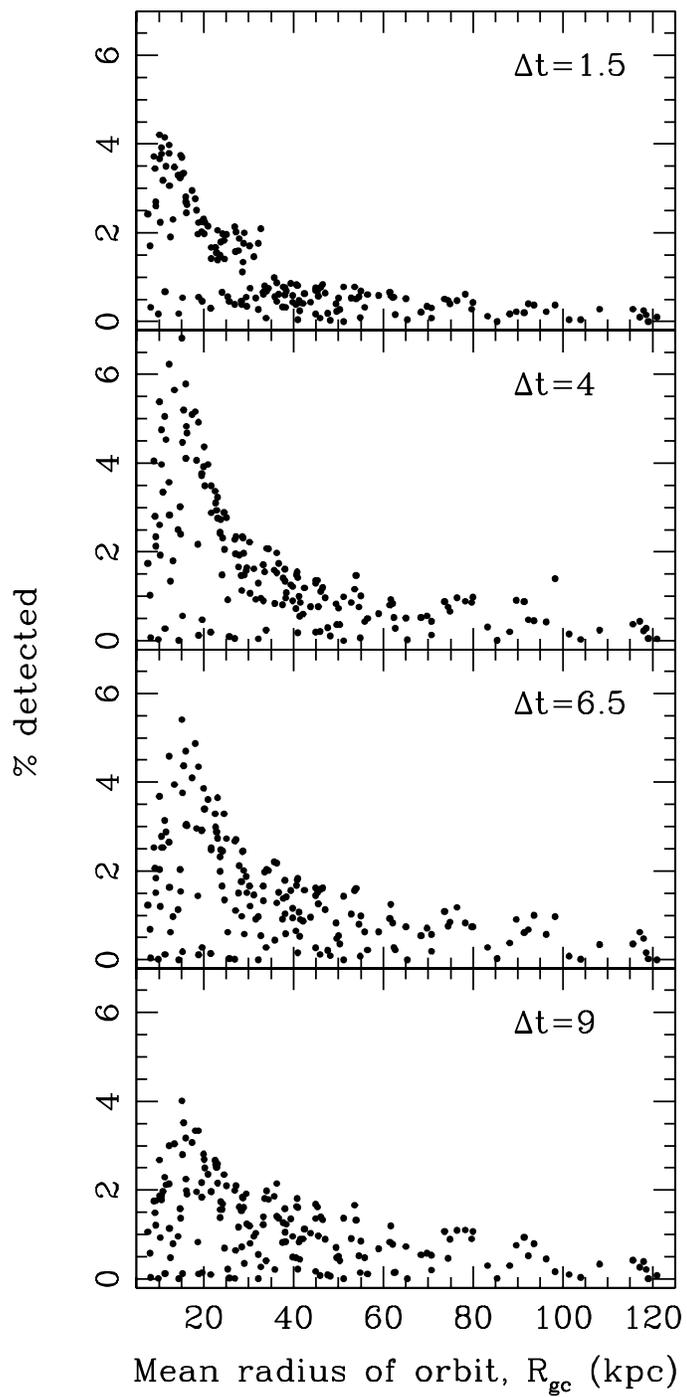
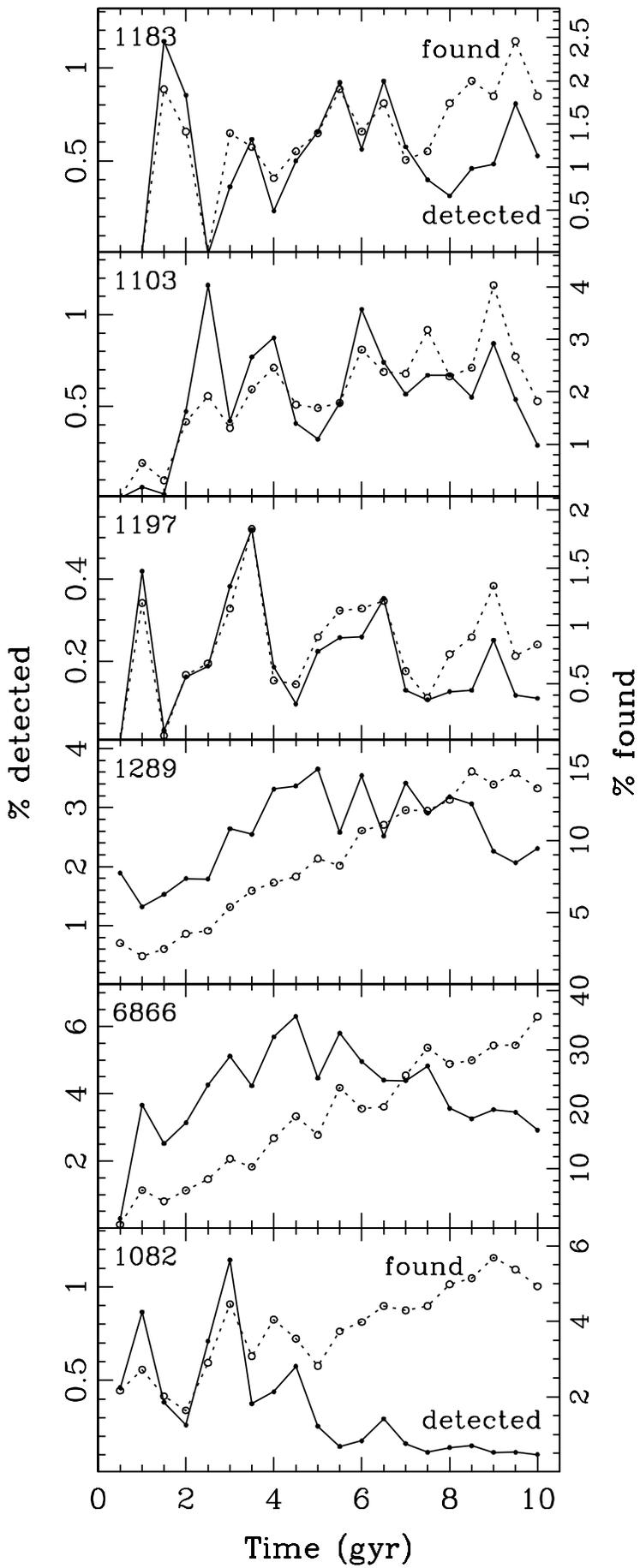
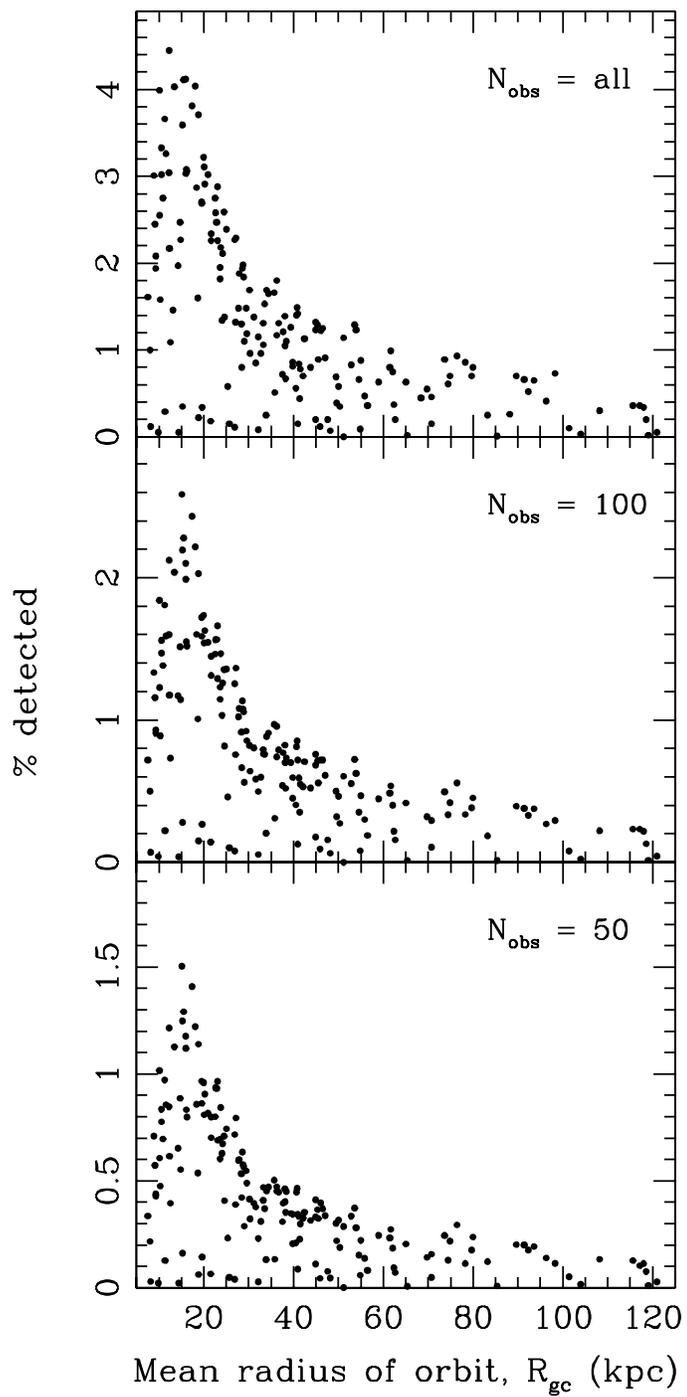
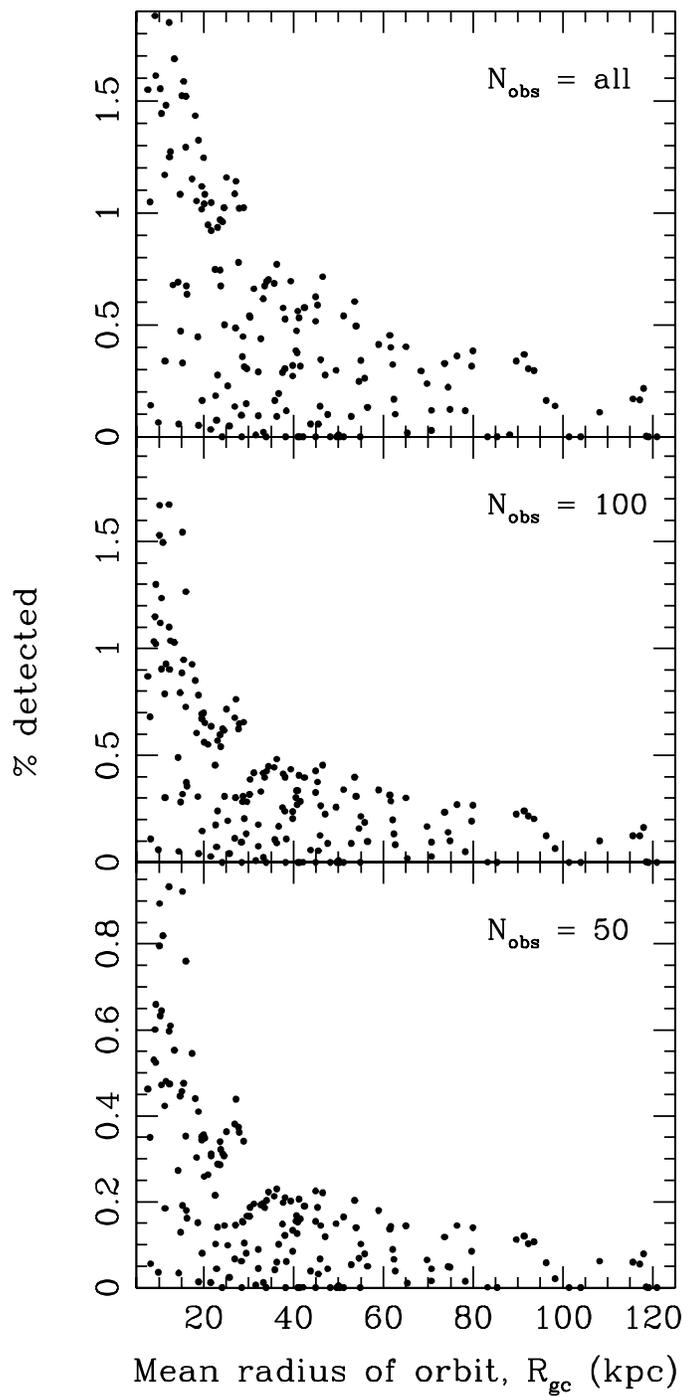
\end{document}